\def\llncs{0}
\def\fullpage{1}
\def\anonymous{0}
\def\draft{1}
\def\submission{0}

\ifnum\submission=1
\def\llncs{1}
\def\draft{1}
\def\anonymous{0}
\fi

\ifnum\llncs=1
    \documentclass{llncs}
\else
\documentclass[a4paper]{article}
    \ifnum\fullpage=1
      \usepackage{fullpage}
      \usepackage[T1]{fontenc}
      \usepackage{stix2}
    \fi
\fi
\usepackage{microtype}
\usepackage{verbatim}
\usepackage{authblk}

\usepackage{graphicx} %
\usepackage{physics} %
\usepackage{xcolor,soul} %
\usepackage{amsmath}
\usepackage{cases}
\usepackage{amssymb}
\DeclareMathAlphabet{\mathpzc}{OT1}{pzc}{m}{it}
\usepackage{mathtools}

\usepackage{amsthm}
\usepackage{enumitem} %
\usepackage{algorithm}
\usepackage{algorithmicx}
\usepackage{algpseudocode}
\usepackage{dsfont}
\usepackage[colorlinks=true,linkcolor=magenta,citecolor=blue,pagebackref=true]{hyperref}
\usepackage{vcell}
\usepackage[flushleft]{threeparttable}
\usepackage{dashbox}
\usepackage{fancybox,framed}
\usepackage[skins]{tcolorbox}
\usepackage{subcaption}
\usepackage[normalem]{ulem}
\usepackage{relsize}
\usepackage[
n,
advantage,
operators,
sets,
adversary,
landau,
probability,
notions,
logic,
ff,
mm,
primitives,
events,
complexity,
asymptotics,
keys,
]{cryptocode}
\usepackage{comment} %
\usepackage[capitalise,nameinlink,noabbrev]{cleveref}
\usepackage{tikz}
\usetikzlibrary{arrows.meta, positioning, shadings}
\usepackage{pifont}
\usepackage{breakcites}

\ifnum\draft=1
	\renewcommand{\comment}[3]{\textcolor{#1}{\(\langle\langle\)#2: #3\(\rangle\rangle\)}}
	\newcommand{\huy}[1]{\comment{purple}{Huy}{#1}}
\else
	\renewcommand{\comment}[3]{}
	\newcommand{\huy}[1]{}
\fi

\ifnum\llncs=0
	\newtheorem{theorem}{Theorem}[section]
	\newtheorem{lemma}[theorem]{Lemma}
	\newtheorem{corollary}[theorem]{Corollary}
	\newtheorem{proposition}[theorem]{Proposition}
	\newtheorem{definition}[theorem]{Definition}

	\newtheorem{construction}[theorem]{Construction}
	\newtheorem{protocol}[theorem]{Protocol}
	\newtheorem{claim}[theorem]{Claim}

	\theoremstyle{remark}

	\newtheorem{conjecture}[theorem]{Conjecture}
\else
	\spnewtheorem{assumption}{Assumption}{\bfseries}{\rmfamily} %
	\spnewtheorem{construction}{Construction}{\bfseries}{\rmfamily} %
\fi
\crefname{appsec}{Appendix}{Appendices}
\crefname{lemma}{Lemma}{Lemmas}
\crefname{definition}{Definition}{Definitions}
\crefname{corollary}{Corollary}{Corollaries}
\crefname{proposition}{Proposition}{Propositions}
\crefname{game}{Game}{Games}
\crefname{fact}{Fact}{Facts}
\crefname{assumption}{Assumption}{Assumptions}
\crefname{construction}{Construction}{Constructions}
\crefname{protocol}{Protocol}{Protocols}

\AddToHook{env/lemma/begin}{\crefalias{theorem}{lemma}}
\AddToHook{env/definition/begin}{\crefalias{theorem}{definition}}
\AddToHook{env/corollary/begin}{\crefalias{theorem}{corollary}}
\AddToHook{env/claim/begin}{\crefalias{theorem}{claim}}
\AddToHook{env/proposition/begin}{\crefalias{theorem}{proposition}}
\AddToHook{env/construction/begin}{\crefalias{theorem}{construction}}
\AddToHook{env/protocol/begin}{\crefalias{theorem}{protocol}}

\crefname{claim}{Claim}{Claims}
\crefname{remark}{Remark}{Remarks}
\crefname{example}{Example}{Examples}
\crefname{note}{Note}{Notes}
\crefname{conjecture}{Conjecture}{Conjectures}

\AddToHook{env/remark/begin}{\crefalias{theorem}{remark}}
\newcommand{\qop}[1]{\mathsf{#1}}

\newcommand{\EE}{\mathop{\mathbb{E}}}

\renewcommand{\epsilon}{\varepsilon}

\renewcommand\prob[2][]{\Pr_{#1}\left[#2\right]}

\newcommand\given[1][]{\:#1\middle\rvert\:}

\renewcommand{\ext}{\mathcal{E}}

\createpseudocodeblock{pcb}{boxed,center}{}{}{} %

\newcommand{\reg}[1]{\mathcal{#1}}
\newcommand{\proj}[1]{\if\relax\detokenize{#1}\relax\Pi\else\Pi_{\mathsf{#1}}\fi}
\newcommand{\ident}{\qop{I}}
\newcommand{\binmeas}[1]{(\proj{#1}, \ident - \proj{#1})}
\newcommand{\meas}[1]{\mathsf{#1}}

\newcommand{\Transform}{\mathsf{Transform}}

\newcommand{\qpt}{\mathsf{QPT}}

\newcommand{\size}[1]{\left|#1\right|}

\newcommand{\IH}{\mathsf{IH}}
\renewcommand{\output}{\mathsf{output}}
\newcommand{\view}{\mathsf{view}}

\DeclareMathOperator{\image}{image}

\newcommand{\rel}[1]{\mathbb{#1}}

\ifnum\llncs=1
    
    \spnewtheorem{claim}{Claim}{\bfseries}{\itshape}
    \numberwithin{claim}{section}
    \crefname{claim}{Claim}{Claims}

    \spnewtheorem{conjecture}{Conjecture}{\bfseries}{\itshape}
    \crefname{conjecture}{Conjecture}{Conjectures}
\fi

\title{A Post-Quantum Interactive Hashing Theorem} 

\ifnum\llncs=1
    \ifnum\anonymous=1
        \author{}
        \institute{}
      \else
        \author{C\'eline Chevalier\inst{1} \and
                Laurent Holin\inst{2} \and Quoc-Huy Vu\inst{3}}
        \institute{CRED, Université Panthéon-Assas, DIENS, École normale supérieure, PSL University, CNRS, Inria, France\\
        \email{celine.chevalier@ens.fr}
        \and DIENS, École normale supérieure, PSL University, CNRS, Inria, France\\
        \email{laurent.holin@ens.fr}
        \and De Vinci Higher Education, De Vinci Research Center, Paris, France\\
        \email{quoc.huy.vu@ens.fr}}
    \fi
\else
    \ifnum\anonymous=1
        \author{}
    \else
        \author[1]{C\'eline Chevalier}
        \author[2]{Laurent Holin}
        \author[3]{Quoc-Huy Vu}
        \affil[1]{CRED, Université Panthéon-Assas, DIENS, École normale supérieure, PSL University, CNRS, Inria, France\par
          \href{mailto:celine.chevalier@ens.fr}{\texttt{celine.chevalier@ens.fr}}}
        \affil[2]{DIENS, École normale supérieure, PSL University, CNRS, Inria, France\par
          \href{mailto:laurent.holin@ens.fr}{\texttt{laurent.holin@ens.fr}}}
        \affil[3]{De Vinci Higher Education, De Vinci Research Center, Paris, France\par
          \href{mailto:quoc.huy.vu@ens.fr}{\texttt{quoc.huy.vu@ens.fr}}}
    \fi
    \date{}
\fi

\begin{document}

\maketitle
\begin{abstract}
  Interactive hashing, introduced by Ostrovsky, Venkatesan, and
  Yung~\cite{DIMACS:OstVenYun93}, is a building block in many cryptographic
  protocols, including perfectly hiding commitments from one-way
  permutations~\cite{C:NOVY92}.
  Its multi-output generalization is a key ingredient in constructions of
  statistical zero-knowledge arguments for \(\mathsf{NP}\) and statistically hiding
  commitments from arbitrary one-way
  functions~\cite{FOCS:NguOngVad06,STOC:HaiRei07}.
  In this note, we give a post-quantum security proof for this multi-output
  interactive hashing protocol with arbitrary relations.
\end{abstract}

\begin{framed}
\small
\noindent\textbf{AI usage statement.} The authors developed an initial
approach to quantum rewinding for interactive hashing, combining the framework
of Lombardi, Ma, and Spooner~\cite{FOCS:LomMaSpo22} with a new coin-regularization
technique.
The first proof draft based on this approach did not establish the claimed
result.
The authors then developed a second proof approach based on Huang and Kalai's
memoryless-rewinding framework~\cite{FOCS:HuaKal25}. 
This approach establishes binding for efficiently verifiable relations.
The authors supplied ChatGPT 5.6 with these two drafts and asked it to combine
the approaches.
ChatGPT-5.6 then produced the current proof, introducing two new ingredients:
the error decomposition and error recording.
The proof is different from the one the authors had in mind when making this
request.
Interestingly, the generated proof crucially relies on a technical lemma (\cref{lem:ih-two-opening-parseval})
proved by the authors for their first approach, although its use in that
approach led to a fatal circularity in the parameter choices.
ChatGPT-6 was subsequently used for proof review, revision, choosing the
parameters, and writing assistance. 
All mathematical claims and arguments in the final manuscript were independently
verified by the authors, who take full responsibility for the content.
According to the \href{https://www.iacr.org/policy/AIPolicy.html}{IACR AI Policy}, we
declare this work as having \textbf{Substantial AI contribution.}
\end{framed}

\ifnum\submission=1
\else
	\clearpage
	\newpage
	\setcounter{tocdepth}{3}
	\tableofcontents
	\newpage
\fi

\section{Introduction}
\label{sec:introduction}

Interactive hashing allows two parties to select a small set containing one
party's private input, while limiting that party's control over the other
outputs. 
Introduced by Ostrovsky, Venkatesan, and Yung~\cite{DIMACS:OstVenYun93}, it was used
by Naor, Ostrovsky, Venkatesan, and Yung to construct perfectly hiding
commitments from one-way permutations~\cite{JC:NOVY98}. 
Its multi-output generalization is a building block in constructions of
statistical zero-knowledge arguments for \(\mathsf{NP}\) and statistically hiding
commitments from arbitrary one-way functions~\cite{FOCS:NguOngVad06,SICOMP:HNORV09}.

Interactive hashing also has applications in quantum cryptography.
Morimae and Yamakawa~\cite{ITCS:MorYam23} use a coherent execution of the NOVY
protocol to construct remote state preparation secure against classical
adversaries, and hence proofs of quantumness, from classically secure
full-domain trapdoor permutations. 
They explicitly left the security of the underlying NOVY construction against
quantum adversaries as an open problem~\cite{ITCS:MorYam23}.

\paragraph{Our contribution.}
We give a post-quantum security proof for the generalized interactive hashing
protocol with multi-bit outputs and \emph{arbitrary relations}, introduced by
Nguyen, Ong, and Vadhan~\cite{FOCS:NguOngVad06}.

The theorem gives two immediate applications.
First, combining the coherent interactive hashing construction
of~\cite{ITCS:MorYam23} with the framework of~\cite{C:BarKhu25}, we obtain
classical verification of quantum computation from post-quantum secure
full-domain trapdoor permutations. 
This answers an open question of Morimae and Yamakawa~\cite{ITCS:MorYam23}.
\begin{corollary}
  There exists a classical verification for quantum computation protocol for
  every language in \(\mathsf{BQP}\), assuming post-quantum secure full-domain
  trapdoor permutations.
\end{corollary}

Second, we recover the original NOVY commitment construction in the post-quantum
setting.
\begin{corollary}
  There exists a post-quantum secure perfect hiding commitment scheme, assuming
  post-quantum secure one-way permutations.
\end{corollary}

\paragraph{Concurrent and independent works.}
During the preparation of this manuscript, we became aware of two concurrent and
independent works establishing post-quantum security of the NOVY bit-commitment
scheme from post-quantum one-way permutations.
Dupuis et al.~\cite{ARXIV:DLMST26} prove sum-binding and collapse-binding and
use the latter to obtain perfect zero-knowledge classical arguments for
\(\mathsf{NP}\) against quantum adversaries.
Chen, Wang, and Xie~\cite{ARXIV:CheWanXie26} prove quantum sum-binding and
extend the construction to efficiently verifiable one-way puzzles with nearly
uniform puzzle distributions. 
They also give an oracle separation ruling out fully black-box constructions of
quantum-computation classical-communication (QCCC) commitments from general
efficiently verifiable one-way puzzles.

Both works analyze the NOVY protocol, whereas our work considers the
multi-output interactive hashing protocol of Nguyen, Ong, and
Vadhan~\cite{FOCS:NguOngVad06} for arbitrary relations.
Although interactive hashing with efficiently verifiable relations
suffices for the applications considered here and in the two concurrent works,
interactive hashing with arbitrary relations is essential to the constructions
and proofs based on \emph{one-way functions}
in~\cite{FOCS:NguOngVad06,STOC:HaiRei07}.
We therefore expect our interactive hashing theorem to be useful in extending
those constructions to the post-quantum setting.
A complete post-quantum analysis of those constructions is left for future work.

\subsection{Technical overview}
\label{sec:ih-technical-overview}

We give an overview of our proof of post-quantum two-opening binding for
the multi-output interactive hashing protocol of~\cite{FOCS:NguOngVad06}
between a sender and a receiver.
In this protocol, the sender holds a secret string \(y\in\bin^{q}\), and the
parties interact for \(m\coloneqq q-k\) rounds, where \(1\leq k<q\).
In round \(i=0,\ldots,m-1\), the receiver sends a uniformly random hash
\(h_{i}\in0^{i}1\bin^{q-i-1}\), and the honest sender replies with the bit
\(c_{i}=h_{i}(y)\coloneqq\langle h_{i},y\rangle\).
The resulting transcript \(\vartheta\) determines an affine bijection
\begin{equation*}
  C_{\vartheta}:\bin^{k}\xrightarrow{\;\cong\;} \{u\in\bin^{q}:h_{i}(u)=c_{i}\text{ for
    all }0\leq i<m\}.
\end{equation*}
The form of the hashes makes these \(m\) equations linearly independent, leaving
exactly \(2^{k}\) compatible strings.
The honest sender's input is one of these strings, and its private output is an
index \(z\in\bin^{k}\) such that \(C_{\vartheta}(z)=y\).

The two-opening binding property is defined as follows.
Fix an arbitrary relation \(\rel W\subseteq\bin^q\times\mathcal X\), where
\(\mathcal X\) is the witness space, and write
\(\rel W_y\coloneqq\{x\in\mathcal X:(y,x)\in\rel W\}\).
We call an execution of a possibly malicious sender against the honest receiver
a \emph{real execution}.
Suppose that, after a real execution, a malicious quantum sender \(S^{*}\)
produces, in a single joint measurement, a pair
\(((x_{0},z_{0}),(x_{1},z_{1}))\) satisfying
\begin{equation*}
  z_{0}\ne z_{1},\qquad
  x_{0}\in\rel W_{C_{\vartheta}(z_{0})},\qquad
  x_{1}\in\rel W_{C_{\vartheta}(z_{1})}
\end{equation*}
with probability at least \(\varepsilon\).
Then there exists an extractor \(\ext\) that, given a uniformly random challenge
\(Y\leftarrow\bin^{q}\), outputs a witness in \(\rel W_{Y}\) with probability
\(\bigOmega{\varepsilon^{2}/(2^{k}q^{2})}\).

\paragraph{Classical proof.}
We first briefly give an overview of the classical proof
of~\cite{FOCS:NguOngVad06}, which is based on a rewinding reduction and a
combinatorial analysis of the protocol's transcript tree.
A node at level \(i\), for \(0\leq i\leq m\), represents a partial transcript
\begin{equation*}
  N=(h_{0},c_{0},\ldots,h_{i-1},c_{i-1}),
\end{equation*}
and \(L_{i}\) denotes the set of all such nodes.
For \(i<m\), each child has the form \((N,h,c)\), where
\(h\in\mathcal H_{i}\coloneqq0^{i}1\bin^{q-i-1}\) is the
next receiver hash and \(c\in\bin\) is the sender's response.
The strings compatible with the prefix form the set
\begin{equation*}
  U(N)\coloneqq\{y\in\bin^{q}:h_{j}(y)=c_{j}\text{ for all }0\leq j<i\},
  \qquad |U(N)|=2^{q-i}.
\end{equation*}
Given a uniformly random challenge \(Y\), the extractor interacts with a
cheating sender and repeatedly rewinds it to select \emph{compatible} receiver
hash functions \(h\) for which the sender's response satisfies \(c=h(Y)\),
thereby keeping \(Y\) in the affine subspace determined by the transcript. 
A balancing argument controls the efficiency of this steering procedure and the
loss in the sender's double-opening probability.
To avoid excessive conditioning, the extractor does not rewind to find
compatible hash functions for all \(m\) rounds.
Instead, it executes the last \(\bigO{\log(q/\varepsilon)}\) rounds honestly
(that is, there is no rewinding in these rounds).
The analysis shows that, if the sender produces two distinct valid openings with
probability \(\varepsilon\), the extractor obtains a witness for \(Y\) with
probability at least \(2^{-k}\poly[\varepsilon/q]\).
This gives two-opening binding whenever finding a witness for a uniformly
random challenge is sufficiently hard.

\paragraph{Overview of our quantum proof.}
The main idea of our proof is to replace classical rewinding with a quantum
rewinding procedure that can bias a malicious sender toward the challenge \(Y\).
We build on the quantum rewinding framework of~\cite{FOCS:LomMaSpo22}, which
uses the singular-vector transformation techniques of~\cite{STOC:GSLW19}.
However, rewinding may alter the sender's residual quantum state and its
probability of producing two distinct outputs with valid witnesses.
Rather than requiring rewinding to preserve this success probability, we develop
a method to account for the errors introduced by this rewinding process using
recording maps.

Our proof consists of four main ingredients:
\begin{itemize}
  \item \textbf{Coin-regularized quantum rewinding.}
    We introduce an auxiliary coin register to overcome the spectral obstruction
    in directly applying singular-vector transformation.
    This yields an efficient rewinding procedure that preserves compatibility
    with the target \(Y\) on every non-aborting branch.
  \item \textbf{From normalization errors to recording.}
    We compare the rewound execution with the real execution and identify the
    normalization errors arising from their difference.
    Although these errors need not be small individually, an operator Fourier
    argument allows us to realize them, up to a scalar factor, as the Kraus
    operators of an efficient recording map.
    In particular, we show that an error operator can be interpreted as a successful branch operator
    for a recorded label.
  \item \textbf{Target switching.}
    Motivated by this recording interpretation, the extractor randomly chooses
    between a normal strategy with no recording, which rewinds toward \(Y\), and
    target-switching strategies, which apply the recording map at an intermediate
    round to obtain a recorded label \(Z\) and subsequently rewind toward \(Z\).
    Every strategy ends with a fresh honest tail, and the extraction target
    remains the original \(Y\).
  \item \textbf{Error decomposition.}
    An exact hybrid identity expresses each scaled branch operator of a real
    execution as a normal term (i.e., error-free), single-error terms, and a
    double-error remainder.
    The normal and single-error terms correspond to the normal and
    target-switching strategies, respectively.
    Each double-error term corresponds to an experiment with two recordings; on
    the successful outcomes being analyzed, the recorded labels are distinct and
    fixed before the honest tail.
    Fresh receiver randomness exponentially suppresses the probability that both
    labels remain compatible with the final transcript, allowing us to bound the
    remainder and conclude that the extraction strategies collectively achieve
    the desired success probability.
\end{itemize}
We now unpack each one.

\paragraph{Coin-regularized quantum rewinding.}
Our starting point is the quantum rewinding framework of \cite{FOCS:LomMaSpo22},
based on the singular-vector transformation techniques of \cite{STOC:GSLW19}
(QSVT). 
The basic idea is to coherently transform the sender's next-message computation
to favor a response compatible with the challenge.

Fix an intermediate node \(N\in L_{i}\), and write \(\qop{K}_{N;h,c}\) for the
sender's next-message Kraus operator corresponding to receiver hash
\(h\in\mathcal H_{i}\) and response \(c\in\{0,1\}\),
including any fresh workspace initialization.
Thus \(\sum_{c}\qop{K}_{N;h,c}^{\dagger}\qop{K}_{N;h,c}=\ident\).
For a target \(y\in U(N)\), define the averaged compatibility operator
\begin{equation*}
  \bar\Pi^{(i)}(y) = \EE_{h\leftarrow\mathcal H_{i}}
  \left[\qop{K}_{N;h,h(y)}^{\dagger} \qop{K}_{N;h,h(y)}\right].
\end{equation*}
This operator measures the sender's average probability of producing a response
\(c\) compatible with \(y\), that is, \(c = h(y)\).
However, directly applying the singular-vector transformation of \cite{STOC:GSLW19},
as used in \cite{FOCS:LomMaSpo22}, encounters a subtle obstruction.
That is, the averaged compatibility operator \(\bar\Pi^{(i)}(y)\), whose
eigenvalues are the squared singular values of the coherent accepting map
constructed from \(\qop{K}_{N;h,h(y)}\), may have zero or arbitrarily small
eigenvalues.
For instance, a malicious sender may, on some quantum state, produce a response
incompatible with \(y\) with probability one.
Consequently, the polar factor of the accepting map may fail to be an isometry
on the entire sender space, while arbitrarily small positive singular values may
render its implementation via QSVT inefficient.

Our idea to overcome this obstruction is to introduce an auxiliary coin
register. 
Fix a small parameter \(\eta\in(0,1/2)\) and prepare the coin in the state
\begin{equation*}
  \ket{\chi_{\eta}}
  =
  \sqrt{1-\eta}\,\ket{0}
  +\sqrt{\eta}\,\ket{1}.
\end{equation*}
Instead of accepting only responses satisfying \(c=h(y)\), the coin-adjusted
compatibility condition is \( c=h(y)\oplus r \), where \(r\) is the coin's
measurement outcome. 
Thus the branch \(r=0\) corresponds to a compatible response, whereas \(r=1\)
corresponds to an incompatible response with a small weight \(\eta\).

We then define the corresponding coherent accepting map as
\begin{align*}
 \qop{B}_{N,y}\ket{\psi}
 =\frac{1}{\sqrt{|\mathcal H_{i}|}}
 \sum_{h\in\mathcal H_{i}}
 \bigl(&
 \sqrt{1-\eta}\,\ket{h,0}
 \qop{K}_{N;h,h(y)}\ket{\psi}\\
 &+\sqrt{\eta}\,\ket{h,1}
 \qop{K}_{N;h,1-h(y)}\ket{\psi}
 \bigr).
\end{align*}
Its associated POVM element \(\qop{Q}_{N,y} \coloneqq \qop{B}_{N,y}^{\dagger} \qop{B}_{N,y}\)
satisfies \(\qop{Q}_{N,y} =\eta \ident+(1-2\eta)\bar\Pi^{(i)}(y)\).
Since \(0\preceq\bar\Pi^{(i)}(y)\preceq \ident\), we obtain
\( \eta \ident\preceq \qop{Q}_{N,y}\preceq(1-\eta)\ident \).
In particular, every singular value of \(\qop{B}_{N,y}\) is at least \(\sqrt{\eta}\),
independently of the malicious sender's strategy.
The coin effectively ensures that this transformation is well-defined and
efficiently implementable.

\paragraph{The rewinding operation.}
We can now define an ideal rewinding step \(\mathsf{Rewind}_{y}\).
Let \(\qop{V}_{N,y} \coloneqq \qop{B}_{N,y}\qop{Q}_{N,y}^{-1/2}\).
Since \(\qop{Q}_{N,y}\succeq\eta \ident\), we have
\(\qop{V}_{N,y}^{\dagger} \qop{V}_{N,y}=\ident \). 
Thus \(\qop{V}_{N, y}\) is an isometry.

At a node \(N\), \(\mathsf{Rewind}_{y}\) applies the isometry \(\qop{V}_{N,y}\) and
measures the hash and coin registers, obtaining \((h,r)\). 
If \(r=1\), it aborts. 
Otherwise, it retains the response \(c=h(y)\) and continues from the child
transcript \((N,h,c)\) with the post-measurement residual sender state.

The non-aborting branch associated with \(h\) is
\begin{equation*}
  \qop{A}_{N;h}^{y}
  =
  \sqrt{\frac{1-\eta}{|\mathcal H_{i}|}}\,
  \qop{K}_{N;h,h(y)}\qop{Q}_{N,y}^{-1/2}.
\end{equation*}
Consequently, every successful rewinding step preserves the compatibility of
\(y\) with the transcript.
We note that the coin does not guarantee that every rewinding step succeeds. 
In particular, a sender may have no compatible-response amplitude on some input
state. 
Its purpose is instead to make the QSVT efficiently implementable, even in such
cases.

To implement this map \(\qop{V}_{N, y}\), we use the singular-vector
transformation in the coherent form recalled in \cref{lem:ih-fixed-transform} from
\cite{FOCS:LomMaSpo22,STOC:GSLW19}. 
\Cref{def:ih-two-opening-polar,lem:ih-two-opening-polar} give the formal definition of \(\mathsf{Rewind}_{y}\) and its efficient approximate
implementation.

\paragraph{The normalization error.}
Although rewinding preserves compatibility with the target, it generally changes
the sender's residual quantum state and the distribution of classical transcript
outcomes \((h, c)\).
Rather than proving that rewinding preserves the sender's success probability
directly, we therefore compare the sender's evolution under rewinding with its
evolution in a real execution, keeping track of both the residual quantum state
and the probability of each outcome.

For a fixed hash function \(h\), write
\(\qop{H}_{N;h} = \frac{1}{\sqrt{|\mathcal H_{i}|}} \qop{K}_{N;h,h(y)}\)
for the corresponding real branch operator.

First, consider the balanced case in which the sender's average probability of
producing a response compatible with a target \(y\) is exactly \(1/2\),
independently of its quantum state.
By its definition, we have
\begin{equation*}
  \bar\Pi^{(i)}(y)=\frac{\ident}{2}.
\end{equation*}
Then \(\qop{Q}_{N,y}=\ident/2\), and the non-aborting rewound branch satisfies
\begin{equation*}
  \qop{A}_{N;h}^{y}=\zeta \qop{H}_{N;h},
  \qquad
  \zeta\coloneqq \sqrt{2(1-\eta)}.
\end{equation*}
Thus, in this balanced case, rewinding amplifies the corresponding real
branch by the fixed factor \(\zeta\).

For a general malicious sender, define the \emph{normalization-error operator}
\begin{equation*}
  \qop{E}_{N,y}\coloneqq \ident-\sqrt{2\qop{Q}_{N,y}}.
\end{equation*}
We then have the exact identity
\begin{equation}
  \label{eq:zetaH_def}
  \zeta \qop{H}_{N;h} = \qop{A}_{N;h}^{y}(\ident-\qop{E}_{N,y}).
\end{equation}
The operator \(\qop{E}_{N,y}\) measures the deviation between the operator-valued
normalization required by the polar transformation and the constant
normalization \(\sqrt{2}\ident\) arising in the balanced case.
It vanishes when \(\bar\Pi^{(i)}(y)=\ident/2\).

Importantly, this normalization error is not an error caused by approximating
QSVT: it is present even for the exact polar isometry. 
Nor can we expect \(\qop{E}_{N,y}\) to have small operator norm for every target. 
A malicious sender may be strongly biased toward or against particular
responses.
Thus, a direct hybrid argument that bounds the normalization errors individually
would not suffice to relate the rewound execution to the real execution.
Instead, we establish a collective operator bound using the Fourier structure of
interactive hashing, as described next.

\paragraph{From normalization errors to recording.}
Define the bias operator
\begin{equation*}
  \qop{D}^{(i)}(y)\coloneqq \bar\Pi^{(i)}(y)-\frac{\ident}{2}.
\end{equation*}
Since \( \qop{Q}_{N,y} = \frac{\ident}{2}+(1-2\eta)\qop{D}^{(i)}(y) \), the normalization
error \(\qop{E}_{N, y}\) is a function of the bias operator \(\qop{D}^{(i)}(y)\).

We prove a technical lemma on the operator Fourier bound (\cref{lem:ih-two-opening-parseval}), which gives
\begin{equation*}
  \sum_{y\in U(N)}\qop{D}^{(i)}(y)^{2}
  =\frac{\ident}{2},
\end{equation*}
which also yields \(\qop{E}_{N,y}^{2}\preceq4\qop{D}^{(i)}(y)^{2}\).
Hence \( \sum_{y\in U(N)}\qop{E}_{N,y}^{2} \preceq2\ident \).

This inequality suggests a different way to handle the normalization errors. 
We can realize each error operator, up to a fixed scalar factor, as the
successful branch operator of a recording map.

Indeed, the operators
\( \qop{R}_{N,y}\coloneqq \frac{\qop{E}_{N,y}}{\sqrt2} \)
satisfy \( \sum_{y\in U(N)}\qop{R}_{N,y}^{\dagger} \qop{R}_{N,y} \preceq \ident \).
Completing these successful Kraus operators with an abort outcome defines a CPTP
map, which we denote by \(\mathsf{Record}_{N}\).
On a successful outcome, the map produces a recorded label \(y\) and applies
the corresponding operator
\(\qop{E}_{N,y}/\sqrt2\) to the residual sender
state.
We note that this recording map \(\mathsf{Record}_{N}\) is efficiently
implementable.
Its formal definition and efficient implementation are given in \cref{def:ih-two-opening-record,lem:ih-two-opening-record}.
Thus a product containing an error operator can be interpreted, up to its scalar
factor, as applying the recording map and continuing with the resulting sender
state.
This is the key observation motivating our extractor.

\paragraph{Why the extractor may switch targets.}
To see how recording leads to extraction, consider a successful real execution
followed by a joint output \(((x_{0},z_{0}),(x_{1},z_{1}))\), with transcript
\(\vartheta\).
Write
\begin{equation*}
  u\coloneqq C_{\vartheta}(z_{0}),\qquad
  v\coloneqq C_{\vartheta}(z_{1}),
  \qquad
  x_{0}\in\rel W_{u},\quad x_{1}\in\rel W_{v}.
\end{equation*}
Here \(z_{0}\ne z_{1}\) are the \(k\)-bit opening indices, while \(u\ne v\) are the
corresponding \(q\)-bit IH output strings, which we call the \emph{endpoints}.

Both \(u\) and \(v\) are compatible with every prefix of this transcript. 
Along this fixed path, write \(\qop{H}_{j}\) for the real branch operator at
round \(j\), and let \(\qop{L}_{j}=\zeta \qop{H}_{j}\).
For \(y\in\{u,v\}\), let \(\qop{A}_{j}^{y} \coloneqq \qop{A}_{N_{j};h_{j}}^{y}\) and
\(\qop{E}_{j}^{y} \coloneqq \qop{E}_{N_{j}, y}\) denote the corresponding rewinding
branch and normalization-error operator, where
\(N_{j} = (h_{0}, c_{0}, \ldots, h_{j-1}, c_{j-1})\) is the transcript's prefix at
round \(j\).
By \cref{eq:zetaH_def}, \( \qop{L}_{j}=\qop{A}_{j}^{y}(\ident-\qop{E}_{j}^{y}) \).
Let
\begin{equation*}
  \qop{A}_{b:a}^{y}=\qop{A}_{b-1}^{y}\cdots \qop{A}_{a}^{y},
  \qquad
  \qop{L}_{b:a}=\qop{L}_{b-1}\cdots \qop{L}_{a}.
\end{equation*}

We first compare the scaled real prefix with rewinding toward \(v\).
For the first \(\alpha\) rounds, where \(\alpha\) is the prefix length chosen
below, a hybrid argument gives
\begin{equation*}
  \qop{L}_{\alpha:0}
  =
  \qop{A}_{\alpha:0}^{v}
  -
  \sum_{j<\alpha}
  \qop{A}_{\alpha:j}^{v}\qop{E}_{j}^{v}\qop{L}_{j:0}.
\end{equation*}
The normal term \(\qop{A}^{v}_{\alpha:0}\) corresponds to rewinding toward \(v\)
throughout the prefix.
This motivates a \emph{normal strategy}, which rewinds toward the externally
chosen target \(Y\).

Each summand contains a normalization-error operator \(\qop{E}_{j}^{v}\).
Since \(\qop{E}_{j}^{v}/\sqrt2\) is the branch
operator for recording \(v\), these summands suggest alternative extraction
strategies.

However, the operator factor \(\qop{L}_{j:0}\) preceding \(\qop{E}_{j}^{v}\) is
still a scaled real prefix.
We therefore apply a second hybrid argument to this prefix, now comparing it
with rewinding toward the other endpoint \(u\):
\begin{equation*}
  \qop{L}_{j:0}
  =
  \qop{A}_{j:0}^{u}
  -
  \sum_{i<j}
  \qop{A}_{j:i}^{u}\qop{E}_{i}^{u}\qop{L}_{i:0}.
\end{equation*}
Substitution gives the exact error decomposition
\begin{equation}
  \label{eq:overview-error-decomposition}
  \begin{aligned}
    \qop{L}_{\alpha:0}
    ={}&
         \qop{A}_{\alpha:0}^{v}
         -
         \sum_{j<\alpha}
         \qop{A}_{\alpha:j}^{v}\qop{E}_{j}^{v}\qop{A}_{j:0}^{u}\\
       &+
         \sum_{0\le i<j<\alpha}
         \qop{A}_{\alpha:j}^{v}\qop{E}_{j}^{v}
         \qop{A}_{j:i}^{u}\qop{E}_{i}^{u}\qop{L}_{i:0}.
  \end{aligned}
\end{equation}
We call \(\qop{A}^{v}_{\alpha:0}\) the \emph{normal term}, the summands in the
first sum \emph{single-error terms}, and those in the second sum
\emph{double-error terms}. 
The sum of all double-error terms is the \emph{double-error remainder}.

After the common honest tail and joint output, the normal term is a branch of
the normal strategy with external target \(Y=v\), returning
\(x_{1}\in\rel W_{v}\).
Each single-error term is \(\sqrt2\) times a
branch of a \emph{target-switching strategy}: rewind toward \(u\), record \(Z=v\) at
round \(j\), and then rewind toward \(v\). 
The external target remains \(Y=u\), so the returned witness is
\(x_{0}\in\rel W_{u}\).
We will analyze and bound the double-error remainder below, after introducing
our extractor, based on this error decomposition.

\paragraph{The extractor.}
The preceding error decomposition motivates an extractor with two types of
execution: the normal strategy and the target-switching strategies.
Formally, we divide the \(m\) rounds into a prefix of length \(\alpha\) and a
fresh honest tail of length \(\beta\). 
For a fixed absolute \(0<\theta<1/4\) and a sufficiently large absolute constant
\(B\), set
\begin{equation*}
  \eta=\frac\theta q,
  \qquad \beta=\min\{m,\lceil\log_{2}(B/\varepsilon)\rceil\},
  \qquad \alpha=m-\beta.
\end{equation*}
The ideal extractor of \cref{con:ih-two-opening-extractor} proceeds as follows,
using ideal rewinding and recording maps and aborting if any invoked procedure
aborts.
\begin{enumerate}
  \item Choose a uniform strategy index \(s\in\{0,\ldots,\alpha\}\).
  \item If \(s=0\), apply \(\mathsf{Rewind}_{Y}\) for all \(\alpha\) prefix
    rounds. 
    If \(j>0\), apply \(\mathsf{Rewind}_{Y}\) for the first \(j-1\) rounds, call
    \(\mathsf{Record}_{N}\) at the current node to obtain \(Z\), and apply
    \(\mathsf{Rewind}_{Z}\) for the remaining prefix rounds.
    Intuitively, index \(0\) selects the normal strategy, and index \(j+1\)
    selects target-switching strategy to record at round \(j\), for
    \(0\leq j<\alpha\).
  \item Run the last \(\beta\) rounds honestly with fresh independent hashes.
    Invoke and measure the sender's joint-output procedure once to obtain
    \(((x_{0},z_{0}),(x_{1},z_{1}))\).
    Return \(x_{b}\) if \(C_{\vartheta}(z_{b})=Y\) and \(z_{0} \neq z_{1}\), for
    some \(b\in\{0,1\}\); otherwise return \(\bot\).
\end{enumerate}

To show that the extractor works, we analyze particular branches of these
strategies.
For the normal strategy, it counts branches in which the external challenge
\(Y\) equals the second IH output string \(v=C_{\vartheta}(z_{1})\), so the extractor
returns \(x_{1}\).
For a target-switching strategy, it counts branches in which the original target
is \(u\) and the recording map outputs \(Z=v\), so the extractor returns \(x_{0}\),
the witness for \(u=C_{\vartheta}(z_{0})\).
 
\paragraph{Why two errors suffice.}
It remains to bound the double-error remainder in \cref{eq:overview-error-decomposition}.
Consider a double-error term with normalization-error operators
\(\qop{E}_{i}^{u}\) and \(\qop{E}_{j}^{v}\) at rounds \(i<j\).
The recording maps have successful Kraus operators
\(\qop{R}_{N_{i},u}=\qop{E}_{i}^{u}/\sqrt{2}\) and
\(\qop{R}_{N_{j},v}=\qop{E}_{j}^{v}/\sqrt{2}\).
Since \(\qop{L}_{i:0}=\zeta^{i} \qop{H}_{i:0}\), we have
\begin{equation*}
  \qop{A}_{\alpha:j}^{v} \qop{E}_{j}^{v} \qop{A}_{j:i}^{u} \qop{E}_{i}^{u} \qop{L}_{i:0}
  =
  2\zeta^{i}
  \qop{A}_{\alpha:j}^{v} \qop{R}_{N_{j},v}
  \qop{A}_{j:i}^{u} \qop{R}_{N_{i},u} \qop{H}_{i:0}.
\end{equation*}
Apart from the scalar factor \(2\zeta^i\), the right-hand side is the prefix Kraus
operator for the branch \((U,V)=(u,v)\) of the following experiment.
It executes the first \(i\) rounds as in a real execution, applies the recording
map to obtain \(U\), rewinds toward \(U\) until round \(j\), applies the recording
map to obtain \(V\), and rewinds toward \(V\) until the end of the rewinding phase.

Both recorded labels \(U,V\) are fixed before the last \(\beta\) rounds.
On the jointly valid outputs counted by this term, they equal the distinct
terminal endpoints \(u,v\), so both must belong to the final interactive-hashing
output set.
We show in \cref{lem:ih-two-opening-tail} that two distinct strings fixed before the honest tail remain
simultaneously compatible after its \(\beta\) rounds with probability at most
\(2^{-\beta}\), even conditioned on the complete preceding history and an arbitrary
residual sender state.
Consequently, each double-error term has norm at most
\begin{equation*}
  2\zeta^{i} 2^{-\beta/2}.
\end{equation*}

This explains why two levels of the error decomposition suffice.
A single-error term need not have small norm, but corresponds to a
target-switching strategy.
Each double-error term instead admits an interpretation involving two recorded
labels, whose simultaneous survival through the honest tail is unlikely.

Moreover, since \(\zeta>1\) is bounded away from one by an absolute constant,
\(\sum_{0\le i<j<\alpha}\zeta^{i} =\bigO{\zeta^{\alpha}}\).
Hence, by the triangle inequality, the double-error remainder has norm at most
\(\bigO{\zeta^{\alpha} 2^{-\beta/2}}\).
If \(\alpha>0\), then \(2^{-\beta}\leq\varepsilon/B\), so choosing \(B\) sufficiently large makes this
bound a small fraction of \(\zeta^{\alpha}\sqrt{\varepsilon}\), the lower bound on the norm of the scaled
real prefix.
If \(\alpha=0\), the double-error remainder is empty.

Applying the triangle inequality to \cref{eq:overview-error-decomposition}, we conclude that the normal and
single-error terms cannot all be small.
These terms correspond to the normal and target-switching strategies,
respectively.
Their interpretations as physical extraction experiments, together with
Cauchy--Schwarz and the random choice of index, yield the ideal extraction bound
\(p_{\mathrm{ext}} =\bigOmega{\frac{\varepsilon^{2}}{2^{k} q^{2}}}\), as established in \cref{prop:ih-two-opening-ideal}.

Finally, replacing the ideal rewinding and recording maps with their
efficient approximate implementations incurs an additive loss of \(\bigO{q\nu}\),
which can be controlled to be at most half the ideal lower bound.
Writing \(\tilde p_{\mathrm{ext}}\) for the extraction probability of the efficient
implementation, we obtain
\begin{equation*}
  \tilde p_{\mathrm{ext}}=\prob{X\in\rel W_{Y}}
  =\bigOmega{\frac{\varepsilon^{2}}{2^{k} q^{2}}},
\end{equation*}
as claimed.

\section{Preliminaries}
We write \(\lambda\) for the security parameter and \(U_d\) for a uniformly
random \(d\)-bit string. 

\paragraph{Relations.}
A relation \(\rel{W}\) is a set of pairs \(({x}, {w})\) where \({x}\) is an
instance and \({w}\) is a witness.
The corresponding language \(\mathcal{L}_{\rel{W}}\) is the set of instances
\({x}\) for which there exists a witness \({w}\) such that
\(({x}, {w}) \in \rel{W}\).

\subsection{Singular vector algorithms}
\label{sec:ih-rewinding-preliminaries}
In this section, we recall a selection of algorithms from \cite{FOCS:LomMaSpo22}
rewinding framework.

\paragraph{Notation.}
Throughout this section we will consider the interaction between two binary
projective measurements \(\meas{A} = \binmeas{A}\), \(\meas{B} = \binmeas{B}\).
Let \(\proj{A}\proj{B} = \qop{V}\Sigma \qop{W}^{\dagger}\) be a fixed singular value
decomposition with singular values \(s_{j}\).
We write \(\ket{v_{j,1}}\) and \(\ket{w_{j,1}}\) for the left and right singular
vectors with \(s_{j} > 0\), and define the Jordan subspace
\(\reg{S}_{j} \coloneqq \mathrm{span}(\ket{v_{j,1}}, \ket{w_{j,1}})\).
When \(s_{j} < 1\) the subspace \(\reg{S}_{j}\) is two-dimensional and also
contains \(\ket{v_{j,0}}, \ket{w_{j,0}}\), which are singular vectors of
\((\ident - \proj{A})(\ident - \proj{B})\) with the same singular value
\(s_{j}\).
Let \(p_{j} \coloneqq s_{j}^{2}\).

\paragraph{Singular-vector transformation.}

\begin{theorem}[{Singular vector transformation
    \cite[Theorem~6.1]{FOCS:LomMaSpo22}}]
  \label{thm:ih-lms-transform}
  There is a uniform family of circuits \( \{\Transform_{a,\delta}\}_{a,\delta\in(0,1]} \)
  with \(\qop{U}_{\meas A}\) and \(\qop{U}_{\meas B}\) gates and of size
  \(O(\log(1/\delta)/\sqrt a)\) such that the
  following holds.
  For a left singular vector \(\ket{v_{j,1}}\) of \(\proj A\proj B\) with singular
  value \(s_{j}\), whenever \(a\leq s_{j}^{2}\), the circuit
  \( \Transform_{a,\delta}[\meas A\to\meas B](\ket{v_{j,1}}) \) returns the corresponding
  right singular vector \(\ket{w_{j,1}}\) with success probability at least \(1-\delta\).
  Each Jordan subspace \(\reg S_{j}\) is invariant under \(\Transform_{a,\delta}\)
  for every \(a\).
\end{theorem}

The next lemma gives a coherent version of \cref{thm:ih-lms-transform}.

\begin{lemma}[Coherent fixed-runtime singular-vector transformation]
  \label{lem:ih-fixed-transform}
  Let \(\meas{A} = \binmeas{A}\), \(\meas{B} = \binmeas{B}\) be projective
  measurements on a common state space \(\reg S\).
  For any threshold \(0<a<1\), define
  \begin{equation*}
    \qop{P}_{a}=\sum_{s_{j}^{2}\geq a}\ketbra{v_{j,1}},
    \qquad
    \qop{T}_{a}=\sum_{s_{j}^{2}\geq a}\ket{w_{j,1}}\bra{v_{j,1}}.
  \end{equation*}
  Here \(\qop{P}_{a}\) projects onto the input singular vectors whose squared
  singular values are at least \(a\).  The partial isometry \(\qop{T}_{a}\)
  specifies the ideal action of \(\Transform_{a}[\meas A\to\meas B]\)
  on this subspace:
  \begin{equation*}
    \qop{T}_{a}\ket{v_{j,1}}=\ket{w_{j,1}}\quad(s_{j}^{2}\geq a),
    \qquad \qop{T}_{a}^{\dagger} \qop{T}_{a}=\qop{P}_{a}.
  \end{equation*}
  In particular, \(\qop{T}_{a}\) is an isometry on \(\image(\qop{P}_{a})\).

  For every \(0<\delta<1\), the transformation admits a coherent efficient implementation
  with auxiliary registers initialized to \(\ket0\). Write
  \begin{equation*}
    \widetilde{\qop{T}}\ket\psi
      =\Transform_{a,\delta}[\meas A\to\meas B](\ket\psi\ket0).
  \end{equation*}
  For every \(\ket\psi\in\image(\qop{P}_{a})\), this implementation satisfies
  \begin{equation*}
    \bigl\|\widetilde{\qop{T}}\ket\psi
          -(\qop{T}_{a}\ket\psi)\ket0\bigr\|
      \leq\delta\|\ket\psi\|.
  \end{equation*}
  The circuit uses \(O(a^{-1/2}\log(2/\delta))\) calls to the coherent
  measurements.
\end{lemma}
\begin{proof}
  Apply \cite[Theorem~26]{STOC:GSLW19} to a SVD
  \(\proj B\proj A=\qop{W}\Sigma \qop{V}^{\dagger}\), with input projector \(\proj A\),
  output projector \(\proj B\), threshold \(\sqrt a\), and projected error
  \(\tau=\delta^{2}/8\).
  Coherent measurement access means
  \(\qop{U}_{\meas D}=\proj D\otimes \qop{X}+(\ident-\proj D)\otimes\ident\), where
  \(D\in\{A,B\}\) and \(\qop{X}\) is the bit-flip gate.
  The phase construction of \cite[Lemma~19]{STOC:GSLW19} computes and uncomputes
  this auxiliary qubit, returning it to \(\ket0\).
  Let \(\qop{C}_{0}\) denote this unitary circuit with exact phase rotations,
  and define its isometry by
  \begin{equation*}
    \qop{W}_{0}\ket\psi=\qop{C}_{0}(\ket\psi\ket0).
  \end{equation*}
  Thus \(\qop{W}_{0}\) includes both the system and auxiliary outputs and satisfies
  \(\qop{W}_{0}^{\dagger} \qop{W}_{0}=\ident\).
  \cite[Theorem~26]{STOC:GSLW19} implies
  \(\bigl|1-(\bra\psi\qop{T}_{a}^{\dagger}\otimes\bra0)
      \qop{W}_{0}\ket\psi\bigr|\leq\tau\) for every
  unit \(\ket\psi\in\image(\qop{P}_{a})\).
  Since both output vectors have norm one,
  \begin{align*}
    \bigl\|\qop{W}_{0}\ket\psi-(\qop{T}_{a}\ket\psi)\ket0\bigr\|^{2}
    &=2-2\operatorname{Re}\bigl[
        (\bra\psi\qop{T}_{a}^{\dagger}\otimes\bra0)
        \qop{W}_{0}\ket\psi\bigr]\\
    &\leq2\tau=\delta^{2}/4,
  \end{align*}
  giving the desired bound.

  With singular-value threshold \(\sqrt a\) and error \(\tau\), the construction
  uses \(d=\bigO{\frac{\log(1/\tau)}{\sqrt a}} =\bigO{a^{-1/2}\log(2/\delta)}\)
  coherent measurement calls, also by \cite[Theorem~26]{STOC:GSLW19}.
\end{proof}

\section{Multi-output interactive hashing}
\begin{definition}\label{def:ih}
  A \emph{multi-output interactive hashing scheme} is a polynomial-time
  protocol \(\langle S_{\IH},R_{\IH}\rangle\) where both parties receive common inputs
  \((1^{q},1^{k})\), \(S_{\IH}\) receives a private input \(y\in\{0,1\}^{q}\), with
  the common output being a circuit \(C:\{0,1\}^{k}\to\{0,1\}^{q}\), and the private
  output of \(S_{\IH}\) being a string \(z\in\{0,1\}^{k}\).
  We denote \(q\) to be the input length and \(k\) to be the output length.
  The protocol \(\langle S_{\IH},R_{\IH}\rangle\) must satisfy the following security
  properties:
  \begin{enumerate}
    \item \emph{(Correctness)} For all \(R^{\ast}\) and all \(y\in\{0,1\}^{q}\),
      letting
      \begin{equation}
        C=\langle S_{\IH}(y),R^{\ast} \rangle(1^{q},1^{k})
        \quad\text{and}\quad
        z=\output_{S_{\IH}}\langle S_{\IH}(y),R^{\ast} \rangle,
      \end{equation}
      we have that \(C(z)=y\).
    \item \emph{(Perfect hiding)} For all \(R^{\ast}\), \((V,Z)\) is distributed
      identically to \((V,U_{k})\), where
      \begin{equation}
        V=\view_{R^{\ast}}\langle S_{\IH}(U_{q}),R^{\ast} \rangle
        \quad\text{and}\quad
        Z=\output_{S_{\IH}}\langle S_{\IH}(U_{q}),R^{\ast} \rangle.
      \end{equation}
    \item \emph{(Post-quantum computational two-opening binding)} Let \(\rel W\) be
      an arbitrary relation, and write \( \rel W_y\coloneqq\{x:(y,x)\in\rel W\} \).
      Let
      \begin{equation}
        C=\langle S^{\ast},R_{\IH} \rangle(1^{q},1^{k})
        \quad\text{and}\quad
        ((x_{0},z_{0}), (x_{1},z_{1}))=\output_{S^{\ast}}\langle S^{\ast},R_{\IH} \rangle,
      \end{equation}
      where \(\output_{S^*}\) denotes one application of the sender's output
      procedure to its residual state, followed by one measurement of the
      designated register containing the two output pairs.
      There exist a constant \(c>0\) and a \(\qpt\) oracle algorithm \(\adv\)
      such that, for every \(\qpt\) sender \(S^*\), every relation \(\rel W\),
      and every inverse-polynomial \(0<\varepsilon\leq1\), the following
      implication holds.
      If
      \begin{equation}
        \Pr[x_{0}\in \rel{W}_{C(z_{0})}\wedge x_{1}\in \rel{W}_{C(z_{1})}\wedge z_{0}\neq z_{1}]>\varepsilon,
      \end{equation}
      where the above probability is over the randomness and measurements of
      \(R_{\IH}\) and \(S^{\ast}\), then
      \begin{equation}
        \Pr_{y\leftarrow\{0,1\}^{q}}\bigl[\adv^{S^{\ast}}(y,1^{q},1^{k},\varepsilon)\in \rel{W}_{y}\bigr]
        >2^{-k}\cdot(\varepsilon/q)^{c}.
      \end{equation}
  \end{enumerate}
\end{definition}

\begin{protocol}[Multi-output interactive hashing
  \(\langle S_{\IH},R_{\IH}\rangle\) \cite{FOCS:NguOngVad06}]
  \label{prot:ih-multi}
  \leavevmode
  \begin{description}
    \item \textbf{Inputs:}
      \begin{enumerate}
        \item Input length \(1^{q}\) and output length \(1^{k}\), both given as
          common input.
        \item String \(y\in\{0,1\}^{q}\), given as private input to sender
          \(S_{\IH}\).
      \end{enumerate}
    \item \textbf{Protocol:} \(R_{\IH}\): Select \(h_{0},h_{1},\ldots,h_{q-k-1}\) such that each
      \(h_{i}\) is a random vector over \(\FF_{2}\) of the form
      \(0^{i}1\{0,1\}^{q-i-1}\) (i.e., \(i\) zeros followed by a one, and random
      choice for the last \(q-i-1\) positions).
      For \(j=0,\ldots,q-k-1\), do the following:
      \begin{align}
        R_{\IH}\to S_{\IH}: & \quad \text{Send }h_{j}.
        \\
        S_{\IH}\to R_{\IH}: & \quad \text{Send }c_{j}=\langle h_{j},y\rangle.
      \end{align}
    \item \textbf{Output:}
      \begin{itemize}
        \item Common output is a circuit \(C:\{0,1\}^{k}\to\{0,1\}^{q}\),
          computing an affine transformation whose image is
          \(\{y:\langle h_{j},y\rangle=c_{j}\ \forall j=0,\ldots,q-k-1\}\).
        \item Output of \(S_{\IH}\) is a string \(z\in\{0,1\}^{k}\) such that
          \(C(z)=y\).
          (In fact, \(z\) can be taken to be the last \(k\) bits of \(y\).)
      \end{itemize}
  \end{description}
\end{protocol}

It is easy to see the correctness of~\cref{prot:ih-multi}.
Its statistical hiding property follows from the classical analysis
in~\cite{FOCS:NguOngVad06}.
In the rest of the paper, we give a post-quantum security proof for the
two-opening binding property.

\section{Post-quantum two-opening binding for arbitrary relations}
\label{sec:ih-two-opening}
\paragraph{Definitions.}
Fix integers \(q\geq2\), \(1\leq k<q\), and let \(m=q-k\).
Fix a relation family \(\mathbb W\coloneqq\{\mathbb W_{y}\}_{y\in\bin^{q}}\). 
Write \(h(y)\coloneqq\langle h,y\rangle\) for \(h, y \in \bin^{q}\). 
We define the tree of transcripts as follows.
\begin{itemize}
  \item For \(0 \leq i < q\), the set of hash functions of the form
    \(0^{i}1\bin^{q-i-1}\) is denoted as \(\mathcal{H}_{i}\), that is
    \(\mathcal{H}_{i} \coloneqq \{0^{i}1w \mid w \in \bin^{q-i-1}\}\).
  \item \textbf{Nodes.}
    A node at level \(i\) is a prefix transcript \(N\) of length \(i\), i.e.,
    \(N=(h_{0},c_{0},\dots,h_{i-1},c_{i-1})\).
    Let \(L_{i}\) denote the set of nodes at level \(i\).
    For \(N\in L_{i}\) define
    \begin{align*}
      U(N) \coloneqq \{y\in\bin^{q}:\ \forall j<i,\ h_{j}(y)=c_{j}\},
      \qquad |U(N)|=2^{q-i}.
    \end{align*}
  \item \textbf{Transcript.}
    A final transcript \(\vartheta\) determines a leaf
    \(N(\vartheta)\in L_{m}\). 
    Its output map is the efficiently computable affine bijection
    \begin{equation*}
      C_{\vartheta}:\bin^{k}\xrightarrow{\;\cong\;}U(N(\vartheta)).
    \end{equation*}
    Thus \(z\in\bin^{k}\) denotes an output index, and \(C_{\vartheta}(z)\)
    denotes the corresponding IH output.
\end{itemize}

Let \(p\) denote the probability, in a real execution followed by the single
measured joint output, of a well-formed, non-aborting output satisfying
\begin{equation}
  z_{0}\ne z_{1},\qquad
  x_{0}\in\rel W_{C_{\vartheta}(z_{0})},\qquad
  x_{1}\in\rel W_{C_{\vartheta}(z_{1})}.
  \label{eq:ih-two-opening-joint}
\end{equation}
We reserve \(\bot\) as a failure symbol outside \(\rel W_{y}\), so
\(\bot\notin\rel W_{y}\) for every \(y\).

\begin{theorem}[Two-opening binding of multi-output interactive hashing]
  \label{thm:ih-two-opening-main}
  The multi-output interactive hashing protocol described in~\cref{prot:ih-multi}
  satisfies post-quantum computational two-opening binding as defined
  in~\cref{def:ih}.
\end{theorem}

\subsection{The extractor}
\subsubsection{Quantum rewinding with the coin trick}

\paragraph{Notation.}
Fix an input string \(y \in \bin^{q}\).
Fix a round \(0\leq i<m\), a node \(N\in L_{i}\).
Let \(\reg S\) be one padded sender space containing all its private, response,
and workspace registers throughout the protocol. Known zero workspaces are
included from the outset; sender unitaries act as the identity on registers
reserved for later rounds. All sender-space operators below are defined on this
whole space, and all operations act as the identity on any inaccessible
reference. Fresh reduction ancillas
are initialized to \(\ket0\) whenever introduced.
\begin{itemize}
  \item Let \(\reg{H}_{i}\) denote a register with basis
    \(\{\ket{h}:h \in \mathcal{H}_{i}\}\), and let
    \(\ket{+_{H}}_{\reg{H}_{i}} \coloneqq \frac{1}{\sqrt{\size{\mathcal{H}_{i}}}}\sum_{h \in \mathcal{H}_{i}} \ket{h}\).
  \item Let \(\qop{S}_{i}^{(h)}\) denote the unitary on \(\reg{S}\) that the
    sender applies at round \(i\) on input a hash function
    \(h \in \mathcal{H}_{i}\), abbreviating \(\qop{S}_{i,N}^{(h)}\) at the fixed
    node.
  \item Let \(\reg{C}_{i}\) denote the subregister of \(\reg{S}\) that the
    sender measures to obtain its response \(c_{i}\) after applying
    \(\qop{S}_{i}^{(h)}\).
  \item Let \(\reg{R}_{i}\) be a single-qubit ancilla register. 
    For \(\eta \in (0, 1 /2)\), let
    \(\ket{\chi_{\eta}}_{\reg{R}_{i}} \coloneqq \sqrt{1-\eta}\ket{0} + \sqrt{\eta}\ket{1}\).
\end{itemize}
Let \(\qop{P}_{c}^{(i)}=\ketbra{c}{c}_{\reg C_{i}}\otimes\ident\), and define
the hash-controlled sender unitary by
\(\qop{T}_{N}=\sum_{h\in\mathcal H_{i}}\ketbra{h}{h} \otimes\ident_{\reg R_{i}}\otimes \qop{S}_{i,N}^{(h)}\).
Define the projector:
\begin{align*}
  \proj{}_{h}^{(i)}(y)
  \coloneqq
  \qop{S}_{i}^{(h)\dagger}\,
  \bigl(\ketbra{\langle h, y\rangle}_{\reg{C}_{i}} \otimes \ident\bigr)\,
  \qop{S}_{i}^{(h)}.
\end{align*}
Intuitively, \(\proj{}_{h}^{(i)}(y)\) projects onto the subspace of \(\reg{S}\)
where the sender's response is compatible with~\(y\) under the hash
function~\(h\) (that is, its output \(c_{i}\) equals \(\langle h, y \rangle\)).

Define the binary projective measurement
\(\meas{C}^{(i)} = (\proj{{C}}^{(i)}, \ident - \proj{{C}}^{(i)})\) where
\begin{align*}
  \proj{C}^{(i)}
    =\sum_{h\in\mathcal H_{i}}\ketbra h_{\reg H_{i}}\otimes
      \Bigl(&\ketbra0_{\reg R_{i}}\otimes\Pi_{h}^{(i)}(y)\\
            &+\ketbra1_{\reg R_{i}}\otimes
                    (\ident-\Pi_{h}^{(i)}(y))\Bigr),
\end{align*}
and \(\meas{U}^{(i)} = (\proj{{U}}^{(i)}, \ident - \proj{{U}}^{(i)})\) where
\begin{align*}
  \proj{U}^{(i)}
    =\ketbra{+_{H}}_{\reg H_{i}}\otimes\ketbra{\chi_{\eta}}_{\reg R_{i}}
       \otimes\ident_{\reg S}.
\end{align*}
Both are projectors on \(\reg H_{i}\otimes\reg R_{i}\otimes\reg S\).
Write \(r_{i}\) for the coin label at round \(i\).
Intuitively, \(r_{i} = 0\) means ``the sender's answer is compatible'',
otherwise \(r_{i} = 1\) means ``the sender's answer is incompatible''.

Define the averaged compatibility operator and the bias operator by
\begin{align*}
  \bar\Pi^{(i)}(y)
  &\coloneqq\frac1{|\mathcal H_{i}|}
    \sum_{h\in\mathcal H_{i}}\Pi_{h}^{(i)}(y),\\
  \qop{D}^{(i)}(y)&\coloneqq\bar\Pi^{(i)}(y)-\frac{\ident}{2}.
\end{align*}
The operator \(\qop{D}^{(i)}(y)\) is the \emph{bias operator}: it measures the
deviation of \(\bar\Pi^{(i)}(y)\) from \(\ident/2\).
We suppress the fixed node \(N\) in this notation. 
For a residual state \(\rho\), its averaged compatibility probability is
\(\Tr[\rho\bar\Pi^{(i)}(y)]\).
The operators \(\Pi_{h}^{(i)}(y)\) are projectors on \(\reg S\), and their
average \(\bar\Pi^{(i)}(y)\) is a positive contraction on the same space.

The next-message Kraus operator for response \(c\) with hash input \(h\), and
its corresponding POVM element, are
\begin{equation*}
  \qop{K}_{N;h,c}=\qop{P}_{c}^{(i)}\qop{S}_{i,N}^{(h)},
  \qquad \qop{M}_{N;h,c}=\qop{K}_{N;h,c}^{\dagger}\qop{K}_{N;h,c}.
\end{equation*}
For fixed \(N,h\), the operators \(\{\qop{M}_{N;h,c}\}_{c\in\bin}\) form a POVM,
since
\begin{equation*}
  \sum_{c\in\bin}\qop{M}_{N;h,c}
  =\sum_{c\in\bin}\qop{K}_{N;h,c}^{\dagger}\qop{K}_{N;h,c}
  =\ident_{\reg S}.
\end{equation*}
The coin-regularized accepting map \(\qop{B}_{N,y}\) prepares
\(\ket{+_{H}}\ket{\chi_{\eta}}\), applies \(\proj C^{(i)}\), and then applies
\(\qop{T}_{N}\). 
Its branch expansion is
\begin{equation}
  \begin{split}
    \qop{B}_{N,y}\ket\psi
    =\frac1{\sqrt{|\mathcal H_{i}|}}\sum_{h\in\mathcal H_{i}}
       \bigl(&\sqrt{1-\eta}\ket{h,0}\qop{K}_{N;h,h(y)}\ket\psi\\
             &+\sqrt\eta\ket{h,1}\qop{K}_{N;h,1-h(y)}\ket\psi\bigr).
  \end{split}
  \label{eq:ih-two-opening-B}
\end{equation}

\paragraph{The coin trick.}
  Using
  \begin{equation*}
    \braket{+_{H}}{h}=\frac{1}{\sqrt{\abs{\mathcal H_{i}}}}, \qquad \braket{\chi_{\eta}}{0}=\sqrt{1-\eta}, \qquad \braket{\chi_{\eta}}{1}=\sqrt{\eta},
  \end{equation*}
  we get
\begin{align}
  \qop{Q}_{N,y}
    &\coloneqq \qop{B}_{N,y}^{\dagger} \qop{B}_{N,y}
     =\frac1{|\mathcal H_{i}|}\sum_{h\in\mathcal H_{i}}
       \bigl((1-\eta)\Pi_{h}^{(i)}(y)
              +\eta(\ident-\Pi_{h}^{(i)}(y))\bigr)\nonumber\\
    &=\eta\ident+(1-2\eta)\bar\Pi^{(i)}(y)
      =\tfrac12\ident+(1-2\eta)\qop{D}^{(i)}(y).
      \label{eq:ih-two-opening-Q}
\end{align}
Consequently,
\begin{equation*}
  \proj U^{(i)}\proj C^{(i)}\proj U^{(i)}
      =\ketbra{+_{H}}{+_{H}}_{\reg H_{i}}\otimes
       \ketbra{\chi_{\eta}}{\chi_{\eta}}_{\reg R_{i}}\otimes\qop{Q}_{N,y}.
\end{equation*}
On \(\image(\proj U^{(i)})\), this product acts as \(\qop{Q}_{N,y}\) on the
sender register while fixing the prepared hash and coin states.
Moreover,
\begin{equation*}
  \eta\ident\preceq\qop{Q}_{N,y}\preceq(1-\eta)\ident.
\end{equation*}
The coin trick gives a lower bound on every singular value on this input space,
independently of the unknown sender state.

\begin{definition}[Ideal rewinding step \(\mathsf{Rewind}\)]
  \label{def:ih-two-opening-polar}
  At a node \(N\in L_{i}\), for \(y\in U(N)\) and the fixed coin parameter
  \(\eta\), the ideal procedure \(\mathsf{Rewind}_{y}\) acts on the supplied
  sender state as follows:
  \begin{enumerate}[leftmargin=*,itemsep=3pt]
    \item \textbf{Initialize.} 
      Prepare the hash register in \(\ket{+_{H}}\) and the coin register in
      \(\ket{\chi_{\eta}}\).
    \item \textbf{Transform.} 
      Apply the ideal action of
      \(\Transform_{\eta}[\meas{U}^{(i)}\to\meas{C}^{(i)}]\) from
      \cref{lem:ih-fixed-transform}.
    \item \textbf{Sender response.} 
      Apply \(\qop{T}_{N}\).
    \item \textbf{Measure and advance.} 
      Measure \((h,r)\).
      If \(r=1\), return \(\bot\). 
      Otherwise retain the response \(c=h(y)\) and the resulting sender state,
      and move to \((N,h,c)\).
  \end{enumerate}
  Before measurement, its isometry on the sender space is
  \begin{equation*}
    \qop{V}_{N,y}=\qop{B}_{N,y}\qop{Q}_{N,y}^{-1/2},
    \qquad \qop{V}_{N,y}^{\dagger} \qop{V}_{N,y}=\ident.
  \end{equation*} 
  Its non-aborting branch operator is
  \begin{equation}
    \qop{A}_{N;h}^{y}
      =\frac{\sqrt{1-\eta}}{\sqrt{|\mathcal H_{i}|}}
        \qop{K}_{N;h,h(y)}\qop{Q}_{N,y}^{-1/2}.
    \label{eq:ih-two-opening-A}
  \end{equation}
\end{definition}

\begin{lemma}[Efficient implementation of \(\mathsf{Rewind}\)]
  \label{lem:ih-two-opening-polar}
  At every intermediate node \(N\in L_{i}\), \(0\leq i<m\), for \(y\in U(N)\)
  and \(0<\nu<1\), the CPTP map of \(\mathsf{Rewind}_{y}\) defined in
  \cref{def:ih-two-opening-polar} can be approximated efficiently with error at
  most \(\nu\), using \(\bigO{\eta^{-1/2}\log(2/\nu)}\) queries to the sender's
  next-message unitaries and their adjoints.
\end{lemma}
\begin{proof}
  Apply \cref{lem:ih-fixed-transform} with \(\meas A=\meas U^{(i)}\),
  \(\meas B=\meas C^{(i)}\), \(a=\eta\), and \(\delta=\nu/2\).
  By the coin trick, every squared singular value on \(\image(\proj U^{(i)})\)
  is at least \(\eta\), so \(\qop{P}_{a}=\proj U^{(i)}\), where \(\qop{P}_{a}\)
  is defined in \cref{lem:ih-fixed-transform}.
  It remains to show that phase operations for the two projectors
  \(\proj U^{(i)}\) and \(\proj C^{(i)}\) can be implemented efficiently with a
  constant number of calls to the sender's unitaries and their adjoints.
  For \(\proj U^{(i)}\), no queries are needed: unprepare the known hash and
  coin states, apply the phase conditional on both registers being zero, and
  prepare the hash and coin states again.
  For \(\proj C^{(i)}\), define the known response projector
  \begin{equation*}
    \qop{M}_{y}=\sum_{h\in\mathcal H_{i},\,r\in\bin} \ket{h,r}\!\bra{h,r}\otimes \qop{P}_{h(y)\oplus r}^{(i)}.
  \end{equation*}
  Since \(\proj C^{(i)}=\qop{T}_{N}^{\dagger} \qop{M}_{y}\qop{T}_{N}\),
  each phase operation for \(\proj C^{(i)}\) requires one call to the sender's
  next-message unitary, a known response phase, and one call to its adjoint.
  
  Thus, the coherent implementation of \(\mathsf{Rewind}_{y}\) approximates
  \(\qop{V}_{N,y}\) within error \(\nu/2\).
  The circuit uses \(\bigO{\eta^{-1/2}\log(2/\nu)}\) oracle calls.
\end{proof}

\subsubsection{From normalization errors to recording}
Next, we prove the following operator Fourier bound.
\begin{lemma}[Operator Fourier bound]
  \label{lem:ih-two-opening-parseval}
  For every intermediate node \(N\in L_{i}\), \(0\leq i<m\),
  \begin{equation}
    \sum_{y\in U(N)}\qop{D}^{(i)}(y)^{2}=\frac{\ident}{2}.
    \label{eq:ih-two-opening-parseval}
  \end{equation}
  Furthermore, there is an efficient CPTP map with a classical outcome, whose
  successful label-\(y\) branch has Kraus operator
  \(\sqrt2\qop{D}^{(i)}(y)\).
\end{lemma}
\begin{proof}
  Recall that every \(h\in\mathcal H_{i}\) has the
  form \( h=0^{i}1w\) where \(w\in\{0,1\}^{q-i-1}\).
  Define
  \begin{equation*}
    \qop{A}_{w}=\qop{S}_{i,N}^{(h)\dagger}
    (\qop{P}_{0}^{(i)}-\qop{P}_{1}^{(i)})\qop{S}_{i,N}^{(h)}.
  \end{equation*}
  The operator \(\qop{A}_{w}\) is a Hermitian unitary, so
  \(\qop{A}_{w}^{2}=\ident\).
  We also write \(\qop{A}_{h}\coloneqq \qop{A}_{w}\) for \(h=0^{i}1w\).
  \(\qop{A}_{h}\) depends on the sender's strategy but \textbf{not} on the
  target~\(y\). 
  The \(y\)-dependence enters only through the character
  \((-1)^{\langle h,y\rangle}\):
  \begin{equation*}
    \Pi_{h}^{(i)}(y)
    =\tfrac12\bigl(\ident+(-1)^{h(y)}\qop{A}_{h}\bigr).
  \end{equation*}
  Squaring the bias operator:
  \begin{equation*}
    \qop{D}^{(i)}(y)^{2}
    \;=\;
    \frac{1}{4|\mathcal H_{i}|^{2}}
    \sum_{h,h'\in\mathcal H_{i}}
    (-1)^{\langle h\oplus h',\,y\rangle}\,\qop{A}_{h} \qop{A}_{h'}.
  \end{equation*}
  Average over \(y\in U(N)\):
  \begin{equation}
    \label{eq:ih-two-opening-avg-D2}
    \EE_{y\in U(N)} \qop{D}^{(i)}(y)^{2}
    \;=\;
    \frac{1}{4|\mathcal H_{i}|^{2}}
    \sum_{h,h'\in\mathcal H_{i}}
    \EE_{y\in U(N)}
    \!\bigl[(-1)^{\langle h\oplus h',\,y\rangle}\bigr]\,
    \qop{A}_{h} \qop{A}_{h'}.
  \end{equation}
\noindent\textbf{Cross terms (\(h\neq h'\)).}
For \(y\in U(N)\), write \(y=(y_{<i},b,t) \), where \(b=y_{i}\in\{0,1\}\) and
\( t=(y_{i+1},\ldots,y_{q-1}) \in\{0,1\}^{q-i-1} \).
We note that for every pair \((b,t)\), there is a unique \(y^{b,t}\in U(N)\)
whose suffix from coordinate \(i\) onward is \((b,t)\), since the first \(i\)
transcript equations uniquely determine the preceding coordinates.
Thus \((b,t)\mapsto y^{b,t}\) is a bijection, and a uniform \(y\in U(N)\) has a
uniform suffix \((b,t)\).
For distinct \(h=0^{i}1w\) and \(h'=0^{i}1w'\), we have \(w\ne w'\) and
\begin{equation*}
  (-1)^{\langle h\oplus h',y^{b,t}\rangle}
    =(-1)^{\langle w\oplus w',t\rangle}.
\end{equation*}
Choose a coordinate where \(w\oplus w'\) is nonzero.
Pairing each \(t\) with the string obtained by flipping that coordinate cancels
the two character values. 
Hence
\begin{equation*}
  \EE_{y\in U(N)}
  \bigl[(-1)^{\langle h\oplus h',\,y\rangle}\bigr]
  = 0.
\end{equation*}

\noindent\textbf{Diagonal terms (\(h=h'\)).}
Here \(h\oplus h=0\), so \((-1)^{\langle 0,y\rangle}=1\), and
\(\qop{A}_{h} \qop{A}_{h} = \qop{A}_{h}^{2} = \ident\).

\noindent\textbf{Result.}
Only diagonal terms survive in~\cref{eq:ih-two-opening-avg-D2}:
\begin{equation*}
  \EE_{y\in U(N)}\qop{D}^{(i)}(y)^{2}
  \;=\;
  \frac{1}{4|\mathcal H_{i}|^{2}}
  \sum_{h\in\mathcal H_{i}}\ident
  \;=\;
  \frac{\ident}{4|\mathcal H_{i}|}.
\end{equation*}
Since \(|U(N)|=2|\mathcal H_{i}|\), this proves
\cref{eq:ih-two-opening-parseval}.

To construct the CPTP map, define
  \begin{equation*}
    \widehat{\qop{A}}_{t}=\frac1{|\mathcal H_{i}|}
    \sum_{w}(-1)^{\langle w,t\rangle}\qop{A}_{w}.
  \end{equation*}
  Then \(\qop{D}^{(i)}(y^{b,t})=(-1)^{b}\widehat{\qop{A}}_{t}/2\).
  For an input sender state \(\ket\psi\), the CPTP map is defined as follows:
  \begin{enumerate}[leftmargin=*,itemsep=3pt]
    \item \textbf{Prepare and reflect.}
      Prepare a uniform \(w\)-register and apply \(\qop{A}_{w}\) controlled by
      that register, obtaining
      \begin{equation*}
        \frac1{\sqrt{|\mathcal H_{i}|}}\sum_{w}\ket w\,\qop{A}_{w}\ket\psi.
      \end{equation*}
    \item \textbf{Fourier transform.}
      Apply Hadamards to the \(w\)-register. 
      The resulting state is
      \begin{equation*}
        \sum_{t}\ket t\,\widehat{\qop{A}}_{t}\ket\psi.
      \end{equation*}
    \item \textbf{Relabel.}
      Append a qubit \(\ket- = \frac{1}{\sqrt2}\sum_{b} (-1)^{b}\ket{b}\), then
      compute \(y^{b, t}\) coherently, giving
      \begin{equation*}
        \frac1{\sqrt2}\sum_{b,t}(-1)^{b}
        \ket{y^{b, t}}\widehat{\qop{A}}_{t}\ket\psi.
      \end{equation*}
    \item \textbf{Measurement.}
      Using \(\qop{D}^{(i)}(y^{b,t})=(-1)^{b}\widehat{\qop{A}}_{t}/2\), the
      resulting isometry is
      \begin{equation}
        \qop{F}_{N}=\sqrt2\sum_{y\in U(N)}\ket y\,\qop{D}^{(i)}(y),
        \qquad \qop{F}_{N}^{\dagger} \qop{F}_{N}=\ident.
        \label{eq:ih-two-opening-fourierisometry}
      \end{equation}
      Measuring \(y\) gives the claimed CPTP map. 
  \end{enumerate}
  Finally, we note that the circuit uses one sender call and one adjoint call.
\end{proof}

\begin{definition}[Ideal recording map \(\mathsf{Record}\)]
  \label{def:ih-two-opening-record}
  Fix an intermediate node \(N\in L_{i}\), \(0\leq i<m\), and the coin parameter
  \(0<\eta<1/2\). 
  For a string \(z\in U(N)\), define the normalization error operator
  \begin{equation*}
    \qop{E}_{N,z}=\ident-\sqrt{2\qop{Q}_{N,z}}.
  \end{equation*}
  The ideal recording map \(\mathsf{Record}_{N}\) takes as input the sender's
  internal state \(\rho\), and acts as follows:
  \begin{enumerate}[leftmargin=*,itemsep=3pt]
    \item \textbf{Measure.}
      Perform the quantum measurement with outcomes \(U(N)\cup\{\bot\}\),
      specified by the Kraus operators \(\qop{R}_{N,z}=\qop{E}_{N,z}/\sqrt2\)
      for \(z\in U(N)\) and
      \(\qop{R}_{N,\bot}=(\ident-\sum_{z}
        \qop{R}_{N,z}^{\dagger}\qop{R}_{N,z})^{1/2}\).
    \item \textbf{Successful output.} 
      On outcome \(Z=z\), return the recorded label \(z\) and retain the
      residual sender state
      \begin{equation*}
        \rho_{z}=\frac{\qop{E}_{N,z}\rho \qop{E}_{N,z}}
        {\Tr[\qop{E}_{N,z}^{2}\rho]},
        \qquad
        \Pr[Z=z]=\frac12\Tr[\qop{E}_{N,z}^{2}\rho].
      \end{equation*}
    \item \textbf{Abort.} 
      On the abort outcome, discard all local registers and return \(\bot\).
  \end{enumerate}
\end{definition}
The following lemma proves that \(\mathsf{Record}_{N}\) defines a CPTP map and
can be approximated efficiently. 
\begin{lemma}[Efficient implementation of \(\mathsf{Record}\)]
  \label{lem:ih-two-opening-record}
  At every intermediate node \(N\in L_{i}\), \(0\leq i<m\), the
  normalization-error operators in \cref{def:ih-two-opening-record} satisfy
  \begin{equation}
    \sum_{y\in U(N)}\qop{E}_{N,y}^{2}\preceq2\ident.
    \label{eq:ih-two-opening-recordbound}
  \end{equation}
  Hence \(\mathsf{Record}_{N}\), including its classical output and abort
  outcome, is a CPTP map.
  Furthermore, for any \(0<\nu<1\), the CPTP map of \(\mathsf{Record}_{N}\) can
  be approximated efficiently with error at most \(\nu\).
\end{lemma}
\begin{proof}
  We first show that \(\mathsf{Record}_{N}\) defines a CPTP map.
  By \cref{claim:boundE},
  \(\qop{E}_{N,y}^{2}\preceq4\qop{D}^{(i)}(y)^{2}\).
  Summing over \(y\in U(N)\) and applying
  \cref{lem:ih-two-opening-parseval} gives \cref{eq:ih-two-opening-recordbound}.

  Next, we implement the recording map \(\mathsf{Record}_N\) in two stages.
  First, \emph{Fourier recording} applies the coherent isometry \(\qop{F}_{N}\)
  from \cref{eq:ih-two-opening-fourierisometry}, whose label-\(z\) branch
  operator is \(\sqrt2\qop{D}^{(i)}(z)\).
  Second, a \emph{conditional correction} applies a quantum operation controlled
  by the label register, with successful Kraus operator \(\qop{G}_{N,z}\)
  constructed as follows.
  Let \(a=1-2\eta\), and define
  \begin{equation*}
    \qop{X}_{N,y}=-2a\qop{D}^{(i)}(y),\qquad
    \qop{G}_{N,y}
      =-a\bigl(\ident+\sqrt{\ident-\qop{X}_{N,y}}\bigr)^{-1}.
  \end{equation*}
  Since \(0\preceq\bar\Pi^{(i)}(y)\preceq\ident\), the bias operator satisfies
  \(-\ident/2\preceq\qop{D}^{(i)}(y)\preceq\ident/2\).
  Hence \(\|\qop{X}_{N,y}\|\leq a<1\) and
  \(\ident-\qop{X}_{N,y}\succeq(1-a)\ident=2\eta\ident\).
  This also yields \( \|\qop{G}_{N,y}\| \leq\frac{a}{1+\sqrt{1-a}}<1 \).
  Recall that \(\ident-\qop{X}_{N,y}=2\qop{Q}_{N,y}\) by
  \cref{eq:ih-two-opening-Q}, and the definition of \(\qop{E}_{N, y}\), we have
  \begin{equation}
    \begin{aligned}
      \qop{G}_{N,y}\sqrt2\qop{D}^{(i)}(y)
        &=\frac1{\sqrt2}
          \bigl(\ident+\sqrt{\ident-\qop{X}_{N,y}}\bigr)^{-1}
          \qop{X}_{N,y}\\
        &=\frac{\ident-\sqrt{\ident-\qop{X}_{N,y}}}{\sqrt2}
          =\frac{\qop{E}_{N,y}}{\sqrt2}.
    \end{aligned}
    \label{eq:ih-two-opening-record-factor}
  \end{equation}
  On failure, discard all local registers and return \(\bot\); on success,
  measure the label register and return its outcome together with the residual
  sender state.
  Thus applying the correction \(\qop{G}_{N,y}\) after a successful Fourier
  recording \(\qop{F}_{N}\) realizes \cref{def:ih-two-opening-record}.

  Finally, we give an efficient approximation to this ideal recording map.
  In fact, it suffices to approximate the correction \(\qop{G}_{N,y}\)
  coherently, controlled by the recorded label \(y\).

  For \(|x|<1\), the binomial series gives
  \begin{equation*}
    \frac1{1+\sqrt{1-x}}
      =\sum_{n\geq1}c_{n}x^{n-1},
    \qquad
    c_{n}\coloneqq\frac{\binom{2n}{n}}{4^{n}(2n-1)}.
  \end{equation*}
  Its coefficients satisfy \(c_{n}\ge0\) and \(\sum_{n\ge1}c_{n}=1\).
  Since \(\norm{\qop{X}_{N,y}}\leq a<1\), the truncated polynomial
  \begin{equation*}
    \qop{G}_{N,y}^{(d)}
      \coloneqq-a\sum_{n=1}^{d}c_{n}\qop{X}_{N,y}^{n-1}
  \end{equation*}
  satisfies
  \begin{equation*}
    \norm{\qop{G}_{N,y}-\qop{G}_{N,y}^{(d)}}
      \leq a\sum_{n>d}c_{n}a^{n-1}
      \leq a^{d+1}.
  \end{equation*}
  As \(a = 1 - 2\eta \leq e^{-2\eta}\), a degree
  \(d=\bigO{\eta^{-1}\log(2/\nu)}\) suffices to make this error at most
  \(\nu/4\).

  We implement this polynomial using the linear-combination and product
  constructions of \cite[Lemmas~52 and~53]{STOC:GSLW19}.
  The Fourier representation in the proof of \cref{lem:ih-two-opening-parseval}
  gives a uniform block encoding of \(2\qop{D}^{(i)}(y)\), and hence of
  \(\qop{X}_{N,y}\).
  By the product construction \cite[Lemma~53]{STOC:GSLW19}, using separate
  ancilla registers gives block encodings of the powers \(\qop{X}_{N,y}^{n-1}\).

  Next, apply the linear-combination construction with blocks
  \(-\qop{X}_{N,y}^{n-1}\) and weights \(ac_n\).
  Since \(\sum_{n=1}^{d}ac_{n}\leq a<1\), assign the remaining weight to a dummy
  zero block.
  The resulting block encoding of \(\qop{G}_{N,y}^{(d)}\) has normalization one
  and requires no amplitude amplification.
  The label dependence enters only through efficiently computable parity phases,
  so the construction is coherent in \(y\).
  A controlled response reflection uses a forward sender call, a controlled
  known response phase, and an adjoint sender call.
  Thus, the block encoding of \(\qop{G}_{N,y}^{(d)}\) uses polynomially many gates and
  sender queries.

  Write the ideal and polynomial successful recording operators as
  \begin{equation*}
    \qop{R}_{N}\coloneqq\sum_{y\in U(N)}\ket y\qop{R}_{N,y},
    \qquad
    \qop{R}_{N}^{(d)}\coloneqq
      \left(\sum_{y\in U(N)}\ketbra{y}{y}
        \otimes\qop{G}_{N,y}^{(d)}\right)\qop{F}_{N}.
  \end{equation*}
  Since \(\qop{F}_{N}\) is an isometry,
  \begin{equation*}
    \norm{\qop{R}_{N}-\qop{R}_{N}^{(d)}}
      \leq\sup_{y\in U(N)}
        \norm{\qop{G}_{N,y}-\qop{G}_{N,y}^{(d)}}
      \leq\nu/4.
  \end{equation*}
  Hence, the approximation error does not depend on the number of labels.

  The coefficients \(c_{n}\) can be computed exactly in polynomial time.
  Let \(T\) be the number of non-oracle elementary gates in the circuit
  implementing \(\qop{R}_{N}^{(d)}\). 
  Implement each gate within operator-norm distance \(\nu/(4T)\) of its ideal
  operation. 
  A hybrid argument then bounds the total operator error by \(\nu/4\).  
  This yields a successful contraction \(\widetilde{\qop{R}}_{N}\) satisfying
  \begin{equation*}
    \norm{\widetilde{\qop{R}}_{N}-\qop{R}_{N}}\leq\nu/2.
  \end{equation*}
  Completing the ideal successful recording contraction \(R_N\) and its
  approximation \(\widetilde R_N\) with abort outcomes, while discarding all
  local registers on abort, yields CPTP maps differing by at most \(\nu\).
  Measuring the recorded label cannot increase this error.
\end{proof}

\begin{claim}
  \label{claim:boundE}
  \(\qop{E}_{N,y}^{2}\preceq4\qop{D}^{(i)}(y)^{2}\).
\end{claim}
\begin{proof}
  Let \(a=1-2\eta\in(0,1)\) and \(\Lambda=\sqrt{\ident+2a\qop{D}^{(i)}(y)}\).
  Since \(\Lambda\succeq0\), we have \(\|(\ident+\Lambda)^{-1}\|\le1\).
  Moreover,
  \begin{equation*}
    (\ident+\Lambda)\qop{E}_{N,y}
    =\ident-\Lambda^{2}
    =-2a\qop{D}^{(i)}(y),
  \end{equation*}
  so
  \begin{equation*}
    \qop{E}_{N,y}=-2a(\ident+\Lambda)^{-1}\qop{D}^{(i)}(y).
  \end{equation*}
  Consequently, for every vector \(\ket\psi\),
  \begin{equation*}
    \|\qop{E}_{N,y}\ket\psi\|
    \le 2a\|\qop{D}^{(i)}(y)\ket\psi\|
    \le 2\|\qop{D}^{(i)}(y)\ket\psi\|.
  \end{equation*}
  Since both operators are Hermitian, this implies
  \(\qop{E}_{N,y}^{2} \preceq 4\qop{D}^{(i)}(y)^{2}\).
\end{proof}

\subsubsection{Description of the extractor}
Fix a constant \(0<\theta<1/4\) and an integer \(B\geq2\) sufficiently large.
Let
\begin{equation}\label{eq:ih-two-opening-parameters}
  \eta=\frac\theta q,\qquad
  \beta=\min\!\left\{m,\left\lceil\log_{2}\frac B\varepsilon\right\rceil\right\},
  \qquad
  \nu=\bigTheta{\frac{\varepsilon^{2}}{2^{k}(q+1)^{3}}}.
\end{equation}
Let \(\alpha=m-\beta\geq0\). 
Note that \(2^{\beta}<2B/\varepsilon\), and \(2^{-\beta}\leq\varepsilon/B\)
whenever \(\alpha>0\).

\begin{construction}[Extractor]\label{con:ih-two-opening-extractor}
  In the following, \(\mathsf{Rewind}\) and \(\mathsf{Record}\) from
  \cref{def:ih-two-opening-polar,def:ih-two-opening-record} are implemented
  efficiently with error at most \(\nu\).
  On input \((\sigma_{\mathsf{in}}^{S^{*}};Y,1^{q},1^{k},\varepsilon)\), the
  extractor \(\ext^{S^{*}}\) uses the parameters in
  \cref{eq:ih-two-opening-parameters}, does the following:
  \begin{enumerate}[leftmargin=*,itemsep=3pt]
    \item Choose a uniform strategy index \(s\in\{0,\ldots,\alpha\}\).
    \item \textbf{Rewound prefix.}
      \begin{itemize}[nosep]
        \item \emph{If \(s=0\):} run \(\mathsf{Rewind}_{Y}\) at rounds
          \(0,\ldots,\alpha-1\).
        \item \emph{If \(s>0\):} let \(j=s-1\) be the recording round.
          Run \(\mathsf{Rewind}_{Y}\) at rounds \(0,\ldots,j-1\); apply
          \(\mathsf{Record}_{N_j}\) to obtain \(Z\); run
          \(\mathsf{Rewind}_{Z}\) at rounds \(j,\ldots,\alpha-1\).
      \end{itemize}
    \item \textbf{Honest continuation.}
      At node \(N \in L_{\alpha}\), choose random
      \(h_{\alpha} \leftarrow \mathcal{H}_{\alpha}, \ldots, h_{\alpha+\beta-1} \leftarrow \mathcal{H}_{\alpha+\beta-1}\),
      and query \(S^{*}\) along the way to arrive at node
      \(\widetilde{N} = (N, h_{\alpha}, c_{\alpha}, \ldots, h_{\alpha+\beta-1}, c_{\alpha+\beta-1}) \in L_{q-k}\).
      These hashes are fresh and independent of the preceding history.
    \item \textbf{Output.}
      Invoke and measure the sender's joint-output procedure to obtain
      \(((x_{0},z_{0}),(x_{1},z_{1}))\).
      Return \(x_{b}\) if \(z_{0} \neq z_{1}\) and \(C_{\vartheta}(z_{b})=Y\)
      for some \(b\in \bin\); otherwise return \(\bot\). 
  \end{enumerate}
\end{construction}

\subsection{Analysis of the extractor}
\subsubsection{The honest tail}

\begin{lemma}[Few bad descendants]
  \label{lem:ih-two-opening-tail}
	Consider a node \(N\in L_{\alpha}\) and two classical strings \(u \neq v\)
  fixed before the last \(\beta\) rounds.
	Consider the continuation of the protocol for \(\beta\) further rounds with an
  honest receiver, resulting in a random descendant leaf
  \(N'\in L_{\alpha+\beta}\).
  Then,
  \begin{equation*}
    \prob{u,v\in U(N')}\le2^{-\beta}.
  \end{equation*}
  The probability includes the receiver's fresh hashes and the sender's
  measurement outcomes.
\end{lemma}
\begin{proof}
  If either string is outside \(U(N)\), the probability is zero.
  Otherwise, let \(d=u\oplus v\neq0\) and consider any round
  \(t\in\{\alpha,\ldots,\alpha+\beta-1\}\), conditioned that both strings are
  still compatible with the current node \(N_{t}\).

  First, we note that the suffix \(d_{\geq t}\) is nonzero.
  Indeed, if it were zero, back-substitution into the equalities
  \(0 = \langle h_{s},d\rangle= d_{s}\oplus\bigoplus_{\ell=s+1}^{q-1}(h_{s})_{\ell} d_{\ell}\)
  for \(s<t\), which follow from compatibility, would force
  \(d_{t-1},\ldots,d_{0}\) to be zero as well, giving \(d=0\).
  Write the fresh hash as \(h_{t}=0^{t}1w\), where \(w\) is uniform.
  Consider two cases:
  \begin{itemize}
    \item If \(d_{>t}\neq0\), then
      \(\langle h_{t},d\rangle=d_{t}\oplus\langle w,d_{>t}\rangle\) is uniform.
    \item If \(d_{>t}=0\), then \(d_{t}=1\), so the two hash values always
    differ.
  \end{itemize}
  Thus, we always have
  \begin{equation*}
    \prob[h_{t}\leftarrow\mathcal H_{t}]{h_{t}(u)=h_{t}(v)}\leq\frac12.
  \end{equation*}
  Both strings can survive this round only if their hash values agree,
  regardless of the sender's response or residual state.
  Applying this conditional bound successively over the \(\beta\) rounds gives
  \(\prob{u,v\in U(N')}\leq2^{-\beta}\).
\end{proof}

\subsubsection{Error decomposition and its operational interpretations}
Denote a complete transcript/output outcome as
\begin{equation*}
  \omega=\bigl(\vartheta,((x_{0},z_{0}),(x_{1},z_{1}))\bigr),
\end{equation*}
where \(\vartheta=(h_{0},c_{0},\ldots,h_{m-1},c_{m-1})\) is the full transcript.
Let \(\mathcal G\) be the set of these outcomes satisfying
\cref{eq:ih-two-opening-joint}, and define their ordered endpoints by
\begin{equation*}
  u\coloneqq C_{\vartheta}(z_{0}),\qquad
  v\coloneqq C_{\vartheta}(z_{1}).
\end{equation*}
Here \(z_{0},z_{1}\in\bin^{k}\) are output indices; the corresponding \(q\)-bit
IH outputs \((u,v)\), opened by the witnesses \((x_{0},x_{1})\), are called the
\emph{endpoints}.
For \(\omega\in\mathcal G\), these endpoints are distinct and compatible with
every prefix of~\(\vartheta\).

Let \(\zeta=\sqrt{2(1-\eta)}\).
At a selected round \(j<\alpha\), let
\begin{equation*}
  \qop{H}_{j}=\frac1{\sqrt{|\mathcal H_{j}|}}\qop{K}_{N_{j};h_{j},c_{j}},
  \qquad \qop{L}_{j}=\zeta \qop{H}_{j}.
\end{equation*}
Thus \(\qop{H}_{j}\) is the real branch operator and \(\qop{L}_{j}\) its scaled
version.
Let \(\qop{A}_{j}^{y}\) denote the branch \cref{eq:ih-two-opening-A} for the
fixed hash and reply, and let \(\qop{E}_{j}^{y}=\qop{E}_{N_{j},y}\). 
For \(y\in \{u,v\}\), compatibility gives the exact factorization
\begin{equation}
  \label{eq:ih-two-opening-factor}
  \qop{L}_{j}=\qop{A}_{j}^{y}\sqrt{2\qop{Q}_{N_{j},y}}
  =\qop{A}_{j}^{y}(\ident-\qop{E}_{j}^{y}).
\end{equation}
Let
\(\qop{A}^{y}_{b:a}=\qop{A}^{y}_{b-1}\cdots \qop{A}^{y}_{a}, \qop{L}_{b:a}=\qop{L}_{b-1}\cdots \qop{L}_{a}\),
and \(\qop{H}_{b:a}=\qop{H}_{b-1}\cdots \qop{H}_{a}\), with an empty product
equal to \(\ident\). 

\begin{lemma}[Error decomposition]
  \label{lem:ih-two-opening-expansion}
  For every such path and ordered pair of compatible endpoints,
  \begin{align}
    \label{eq:ih-two-opening-two}
    \qop{L}_{\alpha:0}
    &= \qop{A}^{v}_{\alpha:0}
      -\sum_{j=0}^{\alpha-1} \qop{A}^{v}_{\alpha:j}\qop{E}_{j}^{v} \qop{A}^{u}_{j:0}\notag\\
    &\quad+\sum_{0\le i<j<\alpha}
      \qop{A}^{v}_{\alpha:j}\qop{E}_{j}^{v} \qop{A}^{u}_{j:i}\qop{E}_{i}^{u} \qop{L}_{i:0}.
  \end{align}
\end{lemma}
\begin{proof}
  For any composable operator sequences \(\qop{X}_{j},\qop{Y}_{j}\), a hybrid
  argument replacing one factor at a time gives
  \begin{equation*}
    \qop{X}_{b:0}-\qop{Y}_{b:0}
    =\sum_{j<b}\qop{Y}_{b:j+1}(\qop{X}_{j}-\qop{Y}_{j})\qop{X}_{j:0}.
  \end{equation*}
  Apply this identity and \cref{eq:ih-two-opening-factor}, first with endpoint
  \(v\), to obtain
  \begin{equation*}
    \qop{L}_{\alpha:0}=\qop{A}^{v}_{\alpha:0}
    -\sum_{j<\alpha}\qop{A}^{v}_{\alpha:j}\qop{E}_{j}^{v} \qop{L}_{j:0}.
  \end{equation*}
  Apply the same identity with endpoint \(u\) to each remaining scaled real
  prefix to obtain:
  \begin{equation*}
    \qop{L}_{j:0}=\qop{A}^{u}_{j:0}-
    \sum_{i<j}\qop{A}^{u}_{j:i}\qop{E}_{i}^{u} \qop{L}_{i:0}.
  \end{equation*}
  The two equalities together prove \cref{eq:ih-two-opening-two}.
\end{proof}
For the rest of the proof, we define the following notation and terminology.
\begin{itemize}
  \item In the error decomposition in \cref{eq:ih-two-opening-two},
    \(\qop{A}^{v}_{\alpha:0}\) is the \emph{normal term}, the summands in the
    first sum are \emph{single-error terms}, and those in the second sum are
    \emph{double-error terms}. 
    The sum of all double-error terms is the \emph{double-error remainder}. 
  \item A \emph{recorded label} is a \(q\)-bit measurement outcome of
    \(\mathsf{Record}\).
    A successful recorded label belongs to \(U(N)\) at the recording node \(N\).
  \item Fix a purification \(\ket\psi\) of the sender's initial input
    \(\sigma_{\mathsf{in}}^{S^{*}}\). 
    For a fixed prefix, let \(\qop{T}_{\omega}\) be the Kraus operator for the
    last \(\beta\) honest rounds and the measured joint output specified by
    \(\omega\).

    For any path-dependent prefix operator \(\qop{P}_{\omega}\), write
    \begin{equation*}
      \norm{\qop{T}\qop{P}}_{\mathcal G}
      \coloneqq\left(\sum_{\omega\in\mathcal G}
        \norm{\qop{T}_{\omega} \qop{P}_{\omega}\ket\psi}^{\,2}\right)^{1/2}.
    \end{equation*}
    We call this the \emph{restricted branch norm}. 
  \item Let \(p_{\mathrm{norm}}\) be the success probability of the normal strategy
    with ideal primitives and no recording, and \(p_{j}\) be that of the
    target-switching strategy with recording at round \(j\), for
    \(0\leq j<\alpha\).
    Both measure the probability of returning a witness for a uniform original
    target \(Y\), conditioned on strategy index \(s=0\) and \(s=j+1\), respectively.
\end{itemize}

We next give an operational interpretation of the terms in
\cref{eq:ih-two-opening-two}. 
The key observation is that the normalization error itself provides a way to
obtain a new rewinding target. 
Although \(\qop{E}_{j}^{y}\) can have negative eigenvalues,
\(\qop{E}_{j}^{y}/\sqrt2\) is the successful branch operator of
\(\mathsf{Record}_{N_{j}}\) for outcome \(y\). 
Thus, up to explicit scalar factors, the normal term and the single-error terms
correspond to successful branches of the ideal extractor, while the double-error
terms correspond to branches of a two-record experiment. 
\begin{lemma}
  \label{lem:ih-two-opening-interpret}
  The restricted branch norms of the terms in \cref{eq:ih-two-opening-two}
  satisfy
  \begin{align}
    \|\qop{T}\qop{L}_{\alpha:0}\|_{\mathcal G}^{\,2}
    &=\zeta^{2\alpha}p,\label{eq:ih-two-opening-honestnorm}\\
    \|\qop{T}\qop{A}^{v}_{\alpha:0}\|_{\mathcal G}^{\,2}
    &\le 2^{q} p_{\mathrm{norm}},\label{eq:ih-two-opening-normalnorm}\\
    \|\qop{T}\qop{A}^{v}_{\alpha:j}\qop{E}_{j}^{v} \qop{A}^{u}_{j:0}\|_{\mathcal G}^{\,2}
    &\le 2^{q+1}p_{j},\label{eq:ih-two-opening-singlenorm}\\
    \|\qop{T}\qop{A}^{v}_{\alpha:j}\qop{E}_{j}^{v} \qop{A}^{u}_{j:i}\qop{E}_{i}^{u} \qop{L}_{i:0}\|_{\mathcal G}^{\,2}
    &\le 4\zeta^{2i}2^{-\beta}\quad(i<j).\label{eq:ih-two-opening-doublenorm}
  \end{align}
\end{lemma}
\begin{proof}
  \emph{The real execution.}
  For \cref{eq:ih-two-opening-honestnorm},
  \(\qop{L}_{\alpha:0}=\zeta^{\alpha} \qop{H}_{\alpha:0}\), so summing
  good-output branch probabilities of the real execution gives
  \(\zeta^{2\alpha}p\).

  \emph{The normal term: no recording.}
  For \cref{eq:ih-two-opening-normalnorm}, consider the ideal extractor without
  recording. 
  For an outcome \(\omega\in\mathcal G\), its branch on input \(Y=v\) rewinds
  toward \(v\) throughout the prefix, then executes the honest continuation and
  returns \(x_{1}\in\rel W_{v}\).
  For uniform \(Y\in\bin^{q}\), let \(X\) be the string returned by this
  strategy. Then
  \begin{align*}
    p_{\mathrm{norm}}=\prob{X\in\rel W_{Y}}
    &\geq\sum_{v\in\bin^{q}}
      \underbrace{\prob{Y=v}}_{2^{-q}}\,
      \prob{\omega\in\mathcal G,\ C_{\vartheta}(z_{1})=v\given Y=v}\\
    &=2^{-q}\sum_{\omega\in\mathcal G}
      \norm{\qop{T}_{\omega}\qop{A}^{v}_{\alpha:0}\ket\psi}^{2}\\
    &=2^{-q}\norm{\qop{T}\qop{A}^{v}_{\alpha:0}}_{\mathcal G}^{2}.
  \end{align*}
  The inequality follows from
  \(\{\omega\in\mathcal G,\ Y=v\}\subseteq\{X\in\rel W_{Y}\}\):
  on this event, the extractor returns \(X=x_{1}\in\rel W_{Y}\).

  \emph{A single-error term: one recording.}
  For \cref{eq:ih-two-opening-singlenorm}, fix the strategy that records at
  round \(j\), and select its branches with \(Y=u\) and recording outcome
  \(Z=v\).
  That is, the extractor first applies \(\mathsf{Rewind}_{u}\) in rounds
  \(0,\ldots,j-1\), records \(v\) at node \(N_{j}\), and applies
  \(\mathsf{Rewind}_{v}\) in rounds \(j,\ldots,\alpha-1\).
  By \cref{def:ih-two-opening-record}, its successful branch is
  \begin{equation*}
    \qop{T}_{\omega} \qop{A}^{v}_{\alpha:j}(\qop{E}_{j}^{v}/\sqrt2)\qop{A}^{u}_{j:0}\ket\psi.
  \end{equation*}
  The recorded label changes the subsequent rewinding target.
  However, the requested witness remains a witness for \(Y=u\). 
  On every counted output, the output pair contains \(x_{0}\in\rel W_{u}\),
  which the extractor returns.
  For uniform \(Y\in\bin^{q}\), let \(X\) be the string returned by this strategy. 
  Then
  \begin{align*}
    p_{j}
    =\prob{X\in\rel W_{Y}}
    &\geq\sum_{u\in\bin^{q}}
      \underbrace{\prob{Y=u}}_{2^{-q}}\,
      \prob{\omega\in\mathcal G,\ C_{\vartheta}(z_{0})=u,
        Z=C_{\vartheta}(z_{1})\given Y=u}\\
    &=2^{-q}\sum_{\omega\in\mathcal G}
      \norm{\qop{T}_{\omega}\qop{A}^{v}_{\alpha:j}
        \frac{\qop{E}_{j}^{v}}{\sqrt2}
        \qop{A}^{u}_{j:0}\ket\psi}^{2}\\
    &=2^{-q-1}
      \norm{\qop{T}\qop{A}^{v}_{\alpha:j}\qop{E}_{j}^{v}
        \qop{A}^{u}_{j:0}}_{\mathcal G}^{2}.
  \end{align*}
  The inequality follows from the fact that
  \(\{\omega\in\mathcal G,\ Y=u,\ Z=v\}\subseteq\{X\in\rel W_{Y}\}\): on this
  event, the extractor returns \(X=x_{0}\in\rel W_{u}=\rel W_{Y}\).
  The condition \(Z=v\) restricts the outcomes of the recording map from
  \cref{def:ih-two-opening-record}.

  \emph{A double-error term: a two-recording experiment.}
  For \cref{eq:ih-two-opening-doublenorm}, consider the following experiment.
  \begin{enumerate}[leftmargin=*]
    \item Run rounds \(0,\ldots,i-1\) honestly, reaching \(N_{i}\).
    \item Apply \(\mathsf{Record}_{N_{i}}\) to obtain \(U\), then run
      \(\mathsf{Rewind}_{U}\) in rounds \(i,\ldots,j-1\).
    \item Apply \(\mathsf{Record}_{N_{j}}\) to obtain \(V\), then run
      \(\mathsf{Rewind}_{V}\) in rounds \(j,\ldots,\alpha-1\).
    \item Run the last \(\beta\) rounds honestly, then invoke and measure the
    sender's joint-output procedure.
    \item Abort on any \(\mathsf{Record}\) or \(\mathsf{Rewind}\) failure. 
  \end{enumerate}
  Its branch with recorded labels \((U,V)=(u,v)\) is
  \begin{equation*}
    \qop{T}_{\omega} \qop{A}^{v}_{\alpha:j}(\qop{E}_{j}^{v}/\sqrt2)
    \qop{A}^{u}_{j:i}(\qop{E}_{i}^{u}/\sqrt2)\qop{H}_{i:0}\ket\psi.
  \end{equation*}
  Let \(N'\) be the leaf reached after the honest continuation.
  Using \(\qop{L}_{i:0}=\zeta^{i}\qop{H}_{i:0}\), we obtain
  \begin{align*}
    &\norm{\qop{T}\qop{A}^{v}_{\alpha:j}\qop{E}_{j}^{v}
      \qop{A}^{u}_{j:i}\qop{E}_{i}^{u}\qop{L}_{i:0}}_{\mathcal G}^{2}\\
    &\quad=4\zeta^{2i}\sum_{\omega\in\mathcal G}
      \norm{\qop{T}_{\omega}\qop{A}^{v}_{\alpha:j}
        \frac{\qop{E}_{j}^{v}}{\sqrt2}\qop{A}^{u}_{j:i}
        \frac{\qop{E}_{i}^{u}}{\sqrt2}\qop{H}_{i:0}\ket\psi}^{2}\\
    &\quad=4\zeta^{2i}
      \prob{\omega\in\mathcal G,\ U=u,\ V=v}\\
    &\quad\leq4\zeta^{2i}
      \sum_{\substack{N\in L_{\alpha}\\a,b\in\bin^{q},\ a\ne b}}
      \prob{N_{\alpha}=N,\ U=a,\ V=b}\,
      \underbrace{\prob{a,b\in U(N')\given N_{\alpha}=N,\ U=a,\ V=b}}
        _{\leq2^{-\beta}}\\
    &\quad\leq4\zeta^{2i}2^{-\beta}.
  \end{align*}
  The first inequality follows from the fact that
  \(\{\omega\in\mathcal G,\ U=u,\ V=v\} \subseteq\{U\ne V,\ U,V\in U(N')\}\).
  The second inequality follows from \cref{lem:ih-two-opening-tail}. 
  This proves \cref{eq:ih-two-opening-doublenorm}.
\end{proof}

\subsubsection{Ideal extraction bound}
We now bound the double-error remainder.
\begin{proposition}[Ideal extraction bound]\label{prop:ih-two-opening-ideal}
  Let \(p_{\mathrm{ext}}\) denote the success probability of the extractor in
  \cref{con:ih-two-opening-extractor}, using ideal implementations of its
  subroutines. 
  If \(p\geq\varepsilon\), then for a sufficiently large fixed \(B\),
  \begin{equation}\label{eq:ih-two-opening-idealbound}
    p_{\mathrm{ext}}
    =\bigOmega{\frac{\varepsilon^{2}}{2^{k}q^{2}}}.
  \end{equation}
\end{proposition}
\begin{proof}
  Applying the triangle inequality to \cref{eq:ih-two-opening-two} and using
  \cref{lem:ih-two-opening-interpret}, we obtain
  \begin{equation}
    \label{eq:ideal-extraction-hybrid}
    \zeta^{\alpha}\sqrt{p}
    \le
    2^{q/2}
    \left(
      \sqrt{p_{\mathrm{norm}}}
      +\sqrt2\sum_{j<\alpha}\sqrt{p_{j}}
    \right)
    +2^{1-\beta/2}
    \sum_{0\le i<j<\alpha}\zeta^{i}.
  \end{equation}

  We first bound the double-error remainder.
  Since \(q\geq2\), \(\eta=\theta/q\), and \(\theta<1/4\), we have
  \(\zeta=\sqrt{2(1-\eta)}\geq\sqrt{2-\theta}>1\).
  Consequently,
  \begin{equation*}
    \begin{aligned}
      \sum_{0\le i<j<\alpha}\zeta^{i}
      &\le
        \zeta^{\alpha}
        \sum_{r=2}^{\infty} (r-1)\zeta^{-r}\\
      &=\frac{\zeta^{\alpha}}{(\zeta-1)^{2}}
        =\bigO{\zeta^{\alpha}}.
    \end{aligned}
  \end{equation*}
  If \(\alpha>0\), our choice of \(\beta\) ensures that
  \(2^{-\beta}\le\varepsilon/B\).
  If \(\alpha=0\), the remainder is empty.
  Thus, by choosing the absolute constant \(B\) sufficiently large, the last
  term in \cref{eq:ideal-extraction-hybrid} is at most
  \(\zeta^{\alpha}\sqrt{\varepsilon}/2\).

  Since \(p\ge\varepsilon\), it follows that
  \begin{equation*}
    \sqrt{p_{\mathrm{norm}}}
    +\sqrt2\sum_{j<\alpha}\sqrt{p_{j}}
    =\bigOmega{2^{-q/2}\zeta^{\alpha}\sqrt{\varepsilon}}.
  \end{equation*}
  By Cauchy--Schwarz,
  \begin{equation*}
    \bigOmega{2^{-q}\zeta^{2\alpha}\varepsilon} = \left(
      \sqrt{p_{\mathrm{norm}}}
      +\sqrt2\sum_{j<\alpha}\sqrt{p_{j}}
    \right)^{2}
    \le
    (2\alpha+1)
    \left(p_{\mathrm{norm}}+\sum_{j<\alpha}p_{j}\right).
  \end{equation*}
  Recall that the ideal extractor chooses each strategy index \(s\) uniformly at random; index \(j+1\) selects recording round \(j\).
  Therefore,
  \begin{equation*}
    \begin{aligned}
      p_{\mathrm{ext}}
      =\frac{p_{\mathrm{norm}}+\sum_{j<\alpha}p_{j}}
        {\alpha+1}
      =\bigOmega{
        \frac{2^{-q}\zeta^{2\alpha}\varepsilon}
        {(\alpha+1)^{2}}
        }
      =\bigOmega{
        \frac{(1-\eta)^{\alpha}\varepsilon}
        {2^{k+\beta}(\alpha+1)^{2}}
        },
    \end{aligned}
  \end{equation*}
  where the last equality follows from \(\zeta^{2}=2(1-\eta)\) and
  \(\alpha+\beta=q-k\).

  Finally, our parameter choices give
  \begin{equation*}
    (1-\eta)^{\alpha}
    \ge 1-\alpha\eta
    \ge 1-\theta,
    \qquad
    2^{-\beta}=\bigOmega{\varepsilon},
    \qquad
    \alpha+1\le q.
  \end{equation*}
  Substituting these bounds yields
  \begin{equation*}
    p_{\mathrm{ext}}
    =\bigOmega{\frac{\varepsilon^{2}}{2^{k}q^{2}}},
  \end{equation*}
  as claimed. 
\end{proof}

\subsubsection{Completion of the binding proof}
\begin{proof}[Proof of \cref{thm:ih-two-opening-main}]
  Assume \(p\geq\varepsilon\).
  Let \(\tilde p_{\mathrm{ext}}\) denote the success probability of the efficient
  implementation of the extractor in \cref{con:ih-two-opening-extractor}.
  By \cref{prop:ih-two-opening-ideal}, the corresponding ideal extraction probability satisfies
  \begin{equation*}
    p_{\mathrm{ext}} =\bigOmega{\frac{\varepsilon^{2}}{2^{k}q^{2}}}.
  \end{equation*}
  By \cref{lem:ih-two-opening-polar,lem:ih-two-opening-record}, each ideal rewinding or recording map can be implemented with
  error at most \(\nu\).
  Since the extractor invokes at most \(q+1\) such operations, a hybrid argument
  and contractivity of trace distance give, using \cref{eq:ih-two-opening-parameters},
  \begin{equation*}
    \abs{\tilde p_{\mathrm{ext}}-p_{\mathrm{ext}}}
    \leq(q+1)\nu
    =\bigO{\frac{\varepsilon^{2}}{2^{k}q^{2}}}.
  \end{equation*}
  Choosing the multiplicative constant in the definition of \(\nu\) sufficiently
  small makes this loss at most half that lower bound. 
  Hence
  \begin{equation*}
    \tilde p_{\mathrm{ext}}
    =\bigOmega{\frac{\varepsilon^{2}}{2^{k}q^{2}}}.
  \end{equation*}
  Moreover, \cref{eq:ih-two-opening-parameters} gives
  \begin{equation*}
    \eta^{-1}=\bigO{q},\qquad
    \log(1/\nu)
    =\bigO{k+\log q+\log(1/\varepsilon)}.
  \end{equation*}
  Since the extractor executes at most \(q\) protocol rounds,
  \cref{lem:ih-two-opening-polar,lem:ih-two-opening-record} imply that its
  worst-case running time is polynomial in the protocol parameters.

  Finally, since \(\varepsilon/q\le 1/2\), the above extraction bound implies
  \begin{equation*}
    \tilde p_{\mathrm{ext}}
    >2^{-k}(\varepsilon/q)^{c}
  \end{equation*}
  for some sufficiently large absolute constant \(c>2\), establishing the
  two-opening binding property of \cref{def:ih}.
\end{proof}

\ifnum\submission=1
\else
	\paragraph{Acknowledgments.}
  We thank the authors of~\cite{ARXIV:CheWanXie26} for discussions on the overlap
  between the two works. 
  This work was supported in part by the French-American ANR-NSF project CTQC
  (ANR-25-CE48-7926-01 CTQC), the French ANR projects SecNISQ (ANR-21-CE47-0014)
  and HQI (ANR-22-PNCQ-0002).
\fi

\newpage

\bibliographystyle{alpha}
\newcommand{\etalchar}[1]{$^{#1}$}

\newpage
\appendix

\end{document}